\documentclass[twocolumn,iop,apj]{aastex63}
\usepackage{amsfonts,amsmath,graphicx,natbib,url,hyperref,nicefrac}
\usepackage{amssymb, verbatim, subfigure, paralist, soul, comment}

\hypersetup{colorlinks=true, urlcolor=blue, citecolor=blue}

\usepackage{color}
\newcommand{\sprout}{\texttt{Sprout}}

\shorttitle{From Type-Ia SNRs to explosion mechanisms} 
\shortauthors{Mandal et al.}

\begin{document}

\title{Deciphering the explosion mechanism of Type-Ia SNe using their remnants I: general properties and a case study on Tycho's SNR}

\author[0000-0001-9484-1262]{Soham Mandal}
\affiliation{Department of Astronomy, University of Virginia, 530 McCormick Road, Charlottesville, VA 22904, USA}
\affiliation{Virginia Institute for Theoretical Astronomy, University of Virginia, Charlottesville, VA 22904, USA}

\author[0000-0003-3638-8943]{N\'uria Torres-Alb\`a}
\affiliation{Department of Astronomy, University of Virginia, 530 McCormick Road, Charlottesville, VA 22904, USA}

\author[0000-0003-3494-343X]{Carles Badenes}
\affiliation{Department of Physics and Astronomy, University of Pittsburgh,
3941 O’Hara Street, Pittsburgh, PA 15260}

\author[0000-0002-1856-9225]{Shazrene Mohamed}
\affiliation{Department of Astronomy, University of Virginia, 530 McCormick Road, Charlottesville, VA 22904, USA}
\affiliation{Virginia Institute for Theoretical Astronomy, University of Virginia, Charlottesville, VA 22904, USA}
\affiliation{South African Astronomical Observatory, P.O Box 9, Observatory, 7935, Cape Town, South Africa}
\affiliation{Department of Astronomy, University of Cape Town, Private Bag X3, Rondebosch, 7701, Cape Town, South Africa}
\affiliation{NITheCS National Institute for Theoretical and Computational Sciences, South Africa}

\email{soham@virginia.edu}

\begin{abstract}

Type-Ia supernovae (SNe), or runaway thermonuclear explosions of white dwarfs (WDs), play a critical role in the chemical evolution of galaxies, and are important cosmological distance indicators due to their ‘standardizable’ lightcurves. Growing evidence, however, suggests greater diversity in their observed lightcurves (and spectra) than thought previously. This is usually attributed to a variety of WD explosion mechanisms and progenitor system properties, but a direct link between the explosion mechanisms and Type-Ia SN observables remains elusive. Here we present a novel approach to identify explosion mechanisms of Type-Ia SNe, by analyzing the sizes of small-scale turbulent substructures of different elements in their extended ejecta, i.e., in Supernova Remnants (SNRs). Our three-dimensional hydrodynamical models show that substructures in an SNR dominated by iron-group elements may have a typical size different from substructures dominated by intermediate mass elements (e.g., Si, S) in the same SNR. This size difference is governed by the explosion mechanism. Applying this approach to Tycho's SNR, we find that its observed structure is most consistent with an SNR model in our suite that originated from a sub-Chandrasekhar mass WD via the double-detonation mechanism. Extending this method to other well-characterized SNRs can let us connect the inferred explosion mechanism to the associated historical SNe, which often have spectra reconstructed through light echo observations.

\end{abstract}


\section{Introduction}  \label{sec:intro}

Supernovae (SNe) are energetic transients resulting from stellar explosions. With the advent of modern high cadence wide-field surveys such as the Palomar Transient Factory \citep[PTF;][]{Law+2009PASP}, the All-Sky Automated Survey for SuperNovae \citep[ASAS-SN;][]{Kochanek+2017PASP}, the Zwicky Transient Facility (ZTF) Bright Transient Survey \citep{Perley+2020ApJ}, the Young Supernova Experiment \citep[YSE;][]{Aleo+2023ApJS}, and the upcoming Legacy Survey of Space and Time \citep[LSST;][]{Ivezic+2019ApJ}, a deluge of SNe discoveries are expected, already surpassing 1000 SNe per year and predicted to go up to 3-4 million SNe per year \citep{LSST2009arXiv}. Given this explosive growth and the already complex classification landscape, it is now more necessary than ever to understand the impact of both intrinsic parameters (e.g., progenitor mass and structure, explosion mechanism) and extrinsic parameters (e.g., interaction with circumstellar medium or CSM, observer line-of-sight effects) on the diversity of SN observables through multidimensional modeling \citep[e.g.,][]{Botyanszki+2018ApJ,Orlando+2024arXiv}. Thermonuclear or Type-Ia SNe are one of the two broad SN subclasses; they happen when  runaway nuclear burning occurs in a degenerate white dwarf (WD), assumed to be rich in carbon and oxygen. They synthesize a major fraction of iron-group elements in the universe and contribute substantially to the production of intermediate-mass elements \citep{Seitenzahl+2017hsn}. They serve as fundamentally important distance indicators in cosmology \citep{Riess+1998AJ,Perlmuttter+1999ApJ}, owing to the famous width-luminosity relation \citep{Rust1974BAAS,Pskovskii1977SvA,Phillips1993ApJ} of their so-called `standardizable' lightcurves.

Type-Ia progenitors were earlier thought to be members of a homogenous WD population with a more or less fixed mass called the Chandrasekhar mass \citep[$\mathrm{M_{ch}}$;][]{Chandra1931ApJ} at explosion. They are now understood to be much more diverse, especially with the discovery of some Type-Ia SNe that are best explained by sub-$\mathrm{M_{ch}}$ \citep[e.g., SN2018byg;][]{De+2019ApJ} or super-$\mathrm{M_{ch}}$ (e.g., SNLS-03D3bb, \cite{Howell+2006Natur}; SN 2006gz, \cite{Hicken+2007ApJ}; SN 2007if, \cite{Scalzo+2010ApJ}) WDs. More recent observations have revealed an increasingly large diversity in Type-Ia SNe lightcurves and spectra \citep{Blondin+2012AJ,Taubenberger2017hsn}, c.f  modern surveys such as the Dark Energy Survey \citep{Abbott+2024ApJ}, which has observed $\sim1500$ Type-Ia SNe to date. This diversity comes, in a large part, from different physical mechanisms or channels that can lead to runaway nuclear burning in a degenerate WD. 

The earliest explosion mechanism suggested was a detonation (supersonically propagating nuclear burning front) in the center of a massive WD, powered by runaway fusion of carbon \citep{Hansen+1969Ap&SS,Arnett1969Ap&SS}. This model had several features that are inconsistent with observed Type-Ia SNe, the most notable one being overproduction of iron \citep{Arnett+1971ApJ}. Since then, many other models have been proposed, from classical deflagration \citep[or subsonically propagating nuclear flame front; see][]{Nomoto+1984ApJ}, to the more recent delayed detonation models \citep{Khokhlov1991A&A}, and the sub-$\mathrm{M_{ch}}$ double detonation models \citep{Sim+2010ApJ,Kromer+2010ApJ}. For a complete discussion on different Type-Ia explosion models and observational features, we refer the reader to the comprehensive review by \cite{Ruiter+2024arXiv}. The general consensus at the time of writing is that no single channel explains all Type-Ia SNe, with growing debate about the likelihood of different explosion channels and WD masses (sub-$\mathrm{M_{ch}}$ vs near-$\mathrm{M_{ch}}$). For example, late-time spectroscopy \citep{Flors+2020MNRAS} suggests the majority of Type-Ia SNe involve sub-$\mathrm{M_{ch}}$ WDs. On the other hand, the presence of stable nickel in nebular spectra of 1986G-like SNe \citep{Kumar+2025arXiv} show that at least a subset of Type-Ia SNe are best explained by near-$\mathrm{M_{ch}}$ models.

Most of our knowledge about the explosion mechanism of observed Type-Ia SNe is inferred from their spectra and lightcurves \citep{Ropke+2012ApJ,Dessart+2014MNRAS,Floers+2019eeu}. Another valuable, but perhaps relatively overlooked avenue for exploring WD explosions are Supernova Remnants (SNRs). Unlike SNe, SNR are extended objects that have expanded for centuries and have radii extending to a few parsecs. They contain signatures of geometry of the exploding stellar ejecta \citep{Ferrand+2021ApJ,Mandal+2023ApJ} and provide accurate estimates of nucleosynthetic yields in the SN since the ejecta becomes optically thin by the remnant phase \citep{Vink2012A&ARv,Vink2017hsn..book}. There are also more SNRs than SNe in our immediate vicinity (the Milky Way and neighboring galaxies), because SNe are rare events \citep[occurring roughly twice per century in a typical spiral galaxy; see][]{Melinder+2012AandA}. Moreover, SNRs probe the immediate surroundings of SNe, i.e. the environment directly shaped by their progenitors \citep{Patnaude+2017ApJ}. In fact, it has been suggested that X-ray spectra of SNRs may be used to discriminate between WD explosion mechanisms \citep{Badenes+2006ApJ}. Therefore, SNRs provide an effective way to study the properties of SNe in the local universe.

\begin{figure}
\centering
\includegraphics[width=0.48\textwidth]{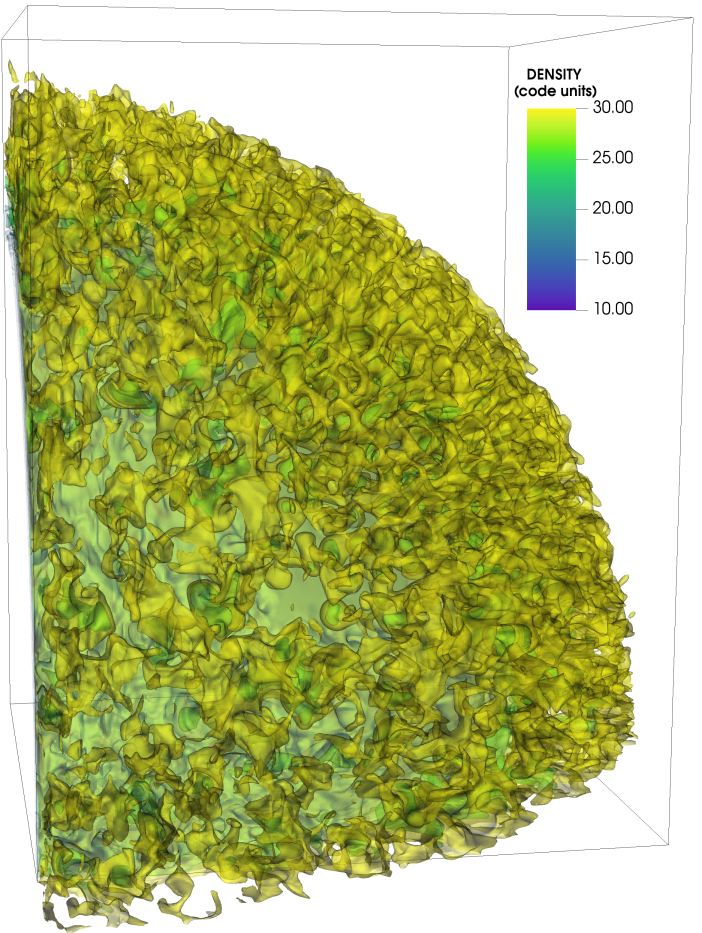}
\caption{Density isosurface in a hydrodynamical SNR model.  The density range shown here focuses on the shocked region between the ejecta and the ambient medium, where turbulence gives rise to small-scale substructures. Rendered using VisIt \citep{HPV:VisIt}.}.
\label{fig:hydro_model}
\end{figure}

\begin{figure*}
\centering
\gridline{\fig{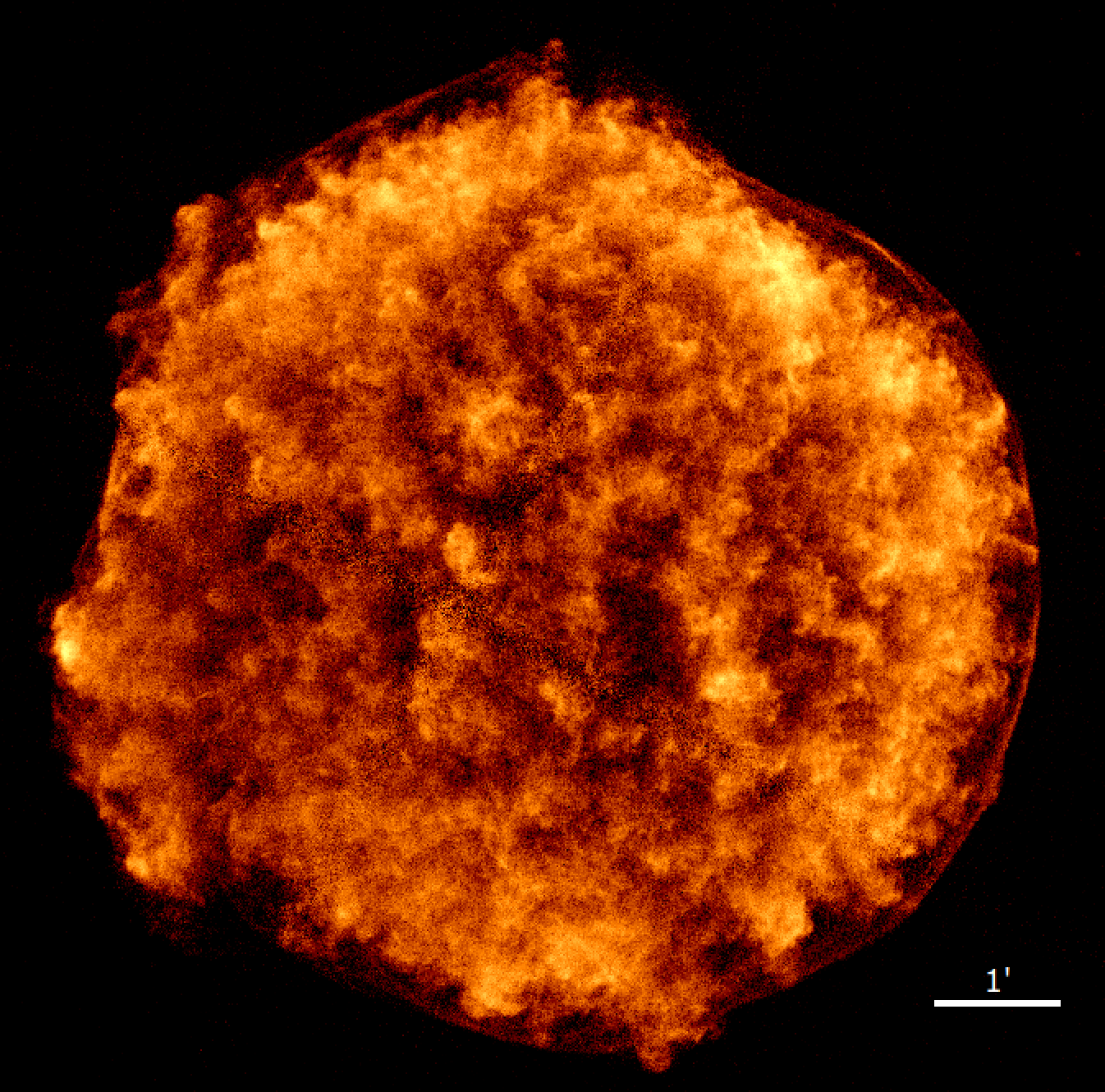}{0.32\textwidth}{}
          \fig{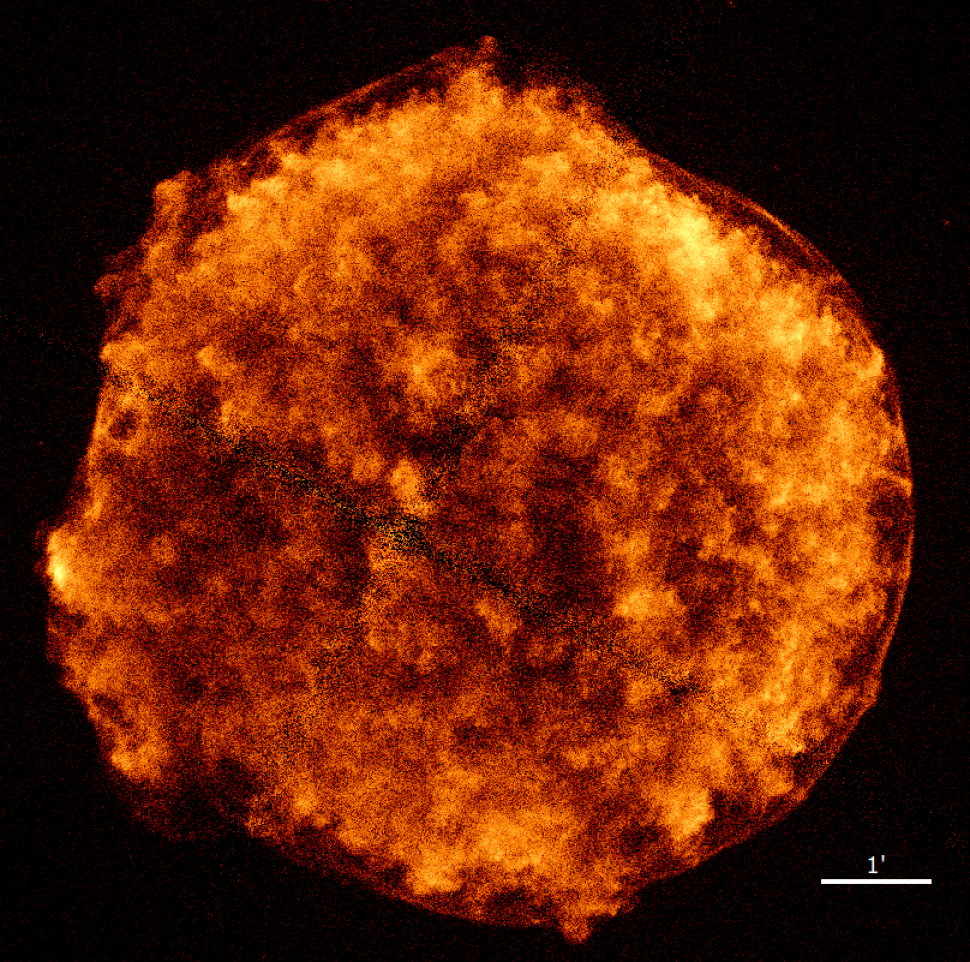}{0.32\textwidth}{}
          \fig{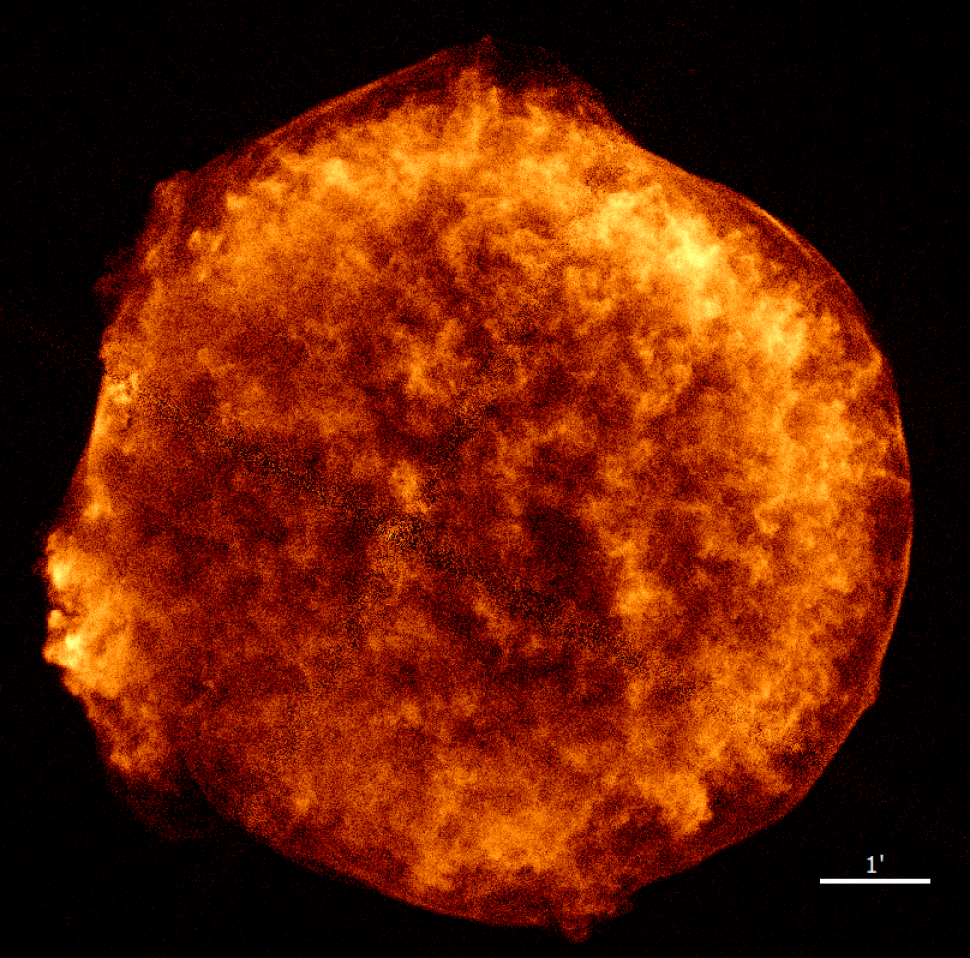}{0.32\textwidth}{}}

\gridline{\fig{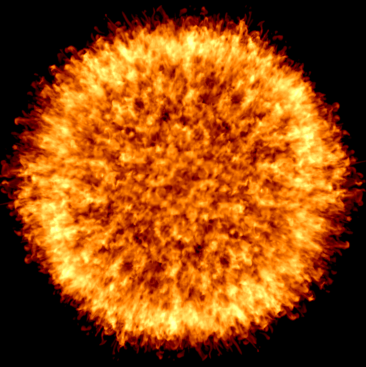}{0.32\textwidth}{}
          \fig{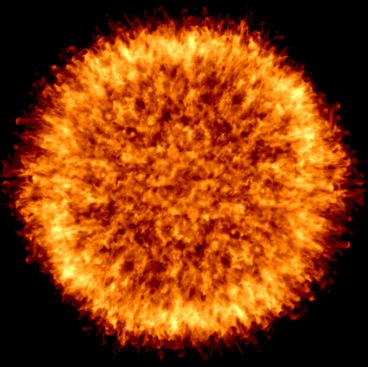}{0.32\textwidth}{}
          \fig{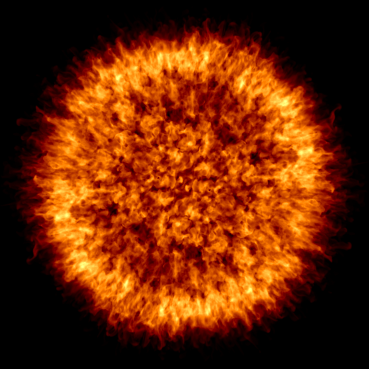}{0.32\textwidth}{}}

\vspace{-8mm}
\caption{\textbf{Top:} Combined \textit{Chandra} images of the Tycho SNR, containing all observations taken in 2009 (and listed in Table~\ref{tab:AllObservations}; see Sect.~\ref{sec:observations} for details). Images have been smoothed with a Gaussian with a radius of three pixels. The images have been filtered to contain only line emission from three elements, from left to right: Si (1.65$-$2.1~keV), S (2.3$-$2.7~keV), and FeL (0.8$-$1.25~keV). \textbf{Bottom:} Synthetic narrowband images corresponding to Si, S and Fe, developed from the DDCO model at a dynamical age $t_D=0.6$ (see Eqn.~\ref{eq:dynamical_age}).}
\label{fig:xray_images}
\end{figure*}

\cite{Polin+2022ApJ} pointed out that the small-scale substructures seen in many SNRs (e.g., Tycho's SNR, SN 1006, Kepler's SNR) have a typical size that depends on the density profile of the SN ejecta. These are formed due to hydrodynamic instabilities (namely, Rayleigh-Taylor Instability or RTI and Kelvin-Helmholtz Instability or KHI) as the ejecta expands into the ambient medium \citep[][also see Fig.~\ref{fig:hydro_model}]{Chevalier+1978ApJ,Chevalier+1992ApJ}. Encouraged by this finding, we speculate whether the substructures dominated by different elements in an SNR should have different typical sizes, linked to their respective density distributions. This could bear the unique signature of the explosion mechanism for the associated SNe. This has already been demonstrated on the observational front by \cite{Lopez+2011ApJ}, who employ the wavelet transform analysis technique to measure the size of substructures of different elements in 24 SNRs. In fact, even visual inspection of narrowband \textit{Chandra} X-ray images of Tycho's SNR \citep[top half of Fig.~\ref{fig:xray_images}, also see][]{Lu+2011ApJ} shows that the substructures dominated by Si, S, and Fe  differ in size.

In this work, we investigate via three-dimensional numerical modeling if one should indeed expect different-sized substructures for different elements in an SNR, and whether this difference can be used to identify the explosion mechanism of the WD. We take Type-Ia SN models with different explosion mechanisms and expand them to the SNR phase. Using power spectra of the SNR models, we measure the typical size of the Si, S, and Fe substructures for all models and compare them. We also generate synthetic narrowband X-ray images from our models to determine if the same information from the 3D models persists when line-of-sight projection effects are taken into account. We apply our approach by examining observations of Tycho's SNR for signs of its explosion mechanism. We use the technique presented in \cite{Mandal+2024ApJ} to analyze both the synthetic images and observations of Tycho's SNR. The SN and SNR models, as well as analysis techniques for both the models and the data are described in Section \ref{sec:numerical}. Details of the observational data used in this work are provided in Section \ref{sec:observations}. The results are discussed in Section \ref{sec:results}, and a discussion and a summary of the results are presented in Sections \ref{sec:discussion} and \ref{sec:conclusion},  respectively.

\section{Numerical method}  \label{sec:numerical}

\subsection{Initial conditions} \label{subsec:initial}



Since the advent of the classical pure detonation \citep{Arnett1969Ap&SS} and pure deflagration \citep{Nomoto+1984ApJ} models, many more scenarios have been proposed to explain the observed features of Type-Ia SNe \citep{Ruiter+2024arXiv}, such as the abundances of iron-group elements (IGEs) and  intermediate mass elements (IMEs) inferred from their spectra. As far as the exploding progenitor is concerned, these models explore two distinct cases:~near-$\mathrm{M_{ch}}$ and sub-$\mathrm{M_{ch}}$ WDs. The most popular example of the former category is the class of ``delayed detonation" models \citep{Khokhlov1991A&A}, where the WD expands due to an initial deflagration phase, followed by a detonation that unbinds the star. In this work, we consider two popular models of this class, namely, the deflagration to detonation transition \citep[DDT;][]{Seitenzahl+2013MNRAS} and the gravitationally confined detonation \citep[GCD;][]{Lach+2022AandA} models. Additionally, we also consider the pure deflagration models of \cite{Fink+2014MNRAS}, which have been found to be good candidates for subluminous 2002cx-like SNe (Type-Iax).

The most popular model for the latter category is the double detonation scenario \citep[DD;][]{Nomoto1982ApJ_2}, where accretion from a companion causes a detonation on the surface of a sub-$\mathrm{M_{ch}}$ WD, leading to its compression and a second detonation inside the WD. In this work, we examine the double detonation of carbon-oxygen (CO) WDs \citep{Gronow+2021AandA} and also oxygen-neon (ONe) WDs. While near-$\mathrm{M_{ch}}$ ONe WDs are not viable SN progenitors (undergoing accretion induced collapse instead), sub-$\mathrm{M_{ch}}$ ONe WDs are found to be suitable candidates for the double detonation mechanism \citep{Marquardt+2015AandA} and can account for a small fraction of Type-Ia SNe. We also examine the case of sub-$\mathrm{M_{ch}}$ WD mergers \citep{Pakmor+2012ApJ}. Although the ejecta mass in this scenario is greater than $\mathrm{M_{ch}}$, this model can be regarded as the pure detonation of a sub-$\mathrm{M_{ch}}$ WD, surrounded by the remains of the secondary WD.

We choose one representative model from the aforementioned groups of models that suitably reproduces observations of known Type-Ia SNe. We use the angle-averaged (1D) versions of these models as our initial conditions. These are hosted on the HESMA project \citep{Kromer+2017MmSAI} website\footnote{\hyperlink{https://hesma.h-its.org}{https://hesma.h-its.org}}. The details of the models used are listed in Table 1. We also plot abundances of Si, S and Fe as a function of Lagrangian mass coordinates for each of these models (used as initial conditions) in Fig.~\ref{fig:model_abundance_profiles}. It can be immediately seen that the near-$\mathrm{M_{ch}}$ models have more Fe near the surface as compared to the sub-$\mathrm{M_{ch}}$ counterparts. This has a very important implication in substructure formation in these SNR models, which we reflect on in Section~\ref{sec:results}.

The angle-averaged initial conditions used in this work remove anisotropies intrinsic to the SN ejecta. However, previous studies that evolve fully three-dimensional explosion models into the remnant phase have shown that explosion-imprinted anisotropies primarily affect the large-scale ($l \lesssim 10$) morphology of the SNR, while intermediate and small-scale structures are dominated by hydrodynamic instabilities that develop during the remnant phase \citep{Ferrand+2019ApJ,Ferrand+2021ApJ}.

In particular, \cite{Mandal+2023ApJ} perform controlled experiments in which small-scale perturbations were seeded in the ejecta and followed through the remnant phase. They show that these perturbations do not persist as distinct features in the power spectrum of the SNR, especially at intermediate and small scales (see their Fig.~6). Instead, within a few decades of SNR evolution, the spectrum becomes dominated by RTI-driven structures (see their Fig. 5). The present study focuses on the statistical properties of small-scale structures in the SNR phase, which typically begins several centuries after the SN explosion. Therefore, we do not expect the omission of explosion-imprinted perturbations in our angle-averaged initial conditions to significantly affect the measured power spectra. In addition, the assumption of spherical symmetry allows us to model only an octant of the remnant \citep[as in][]{Mandal+2023ApJ} without imposing artificial symmetries on the system, thereby increasing the effective spatial resolution for a fixed computational cost.

The models provide us with the ejecta density profile and abundances for Si, S, and Fe. These elements were chosen since they are produced abundantly during SN nucleosynthesis, and have prominent emission lines in soft X-rays. The ejecta is assumed to be cold ($P = 10^{-6}\rho$) and expanding homologously ($v=r/t$). The ambient medium (AM) is taken to be uniform with a density of $n_0 = 0.2\,\mathrm{cm}^{-3}$. This value was chosen to be consistent with the inferred AM density for Tycho's SNR \citep{Cassam-Chenai+2007ApJ,Katsuda+2010ApJ}. These profiles are mapped onto a 3D domain and allowed to expand to the SNR phase, as detailed in the next section.

\begin{deluxetable*}{cccc}
\tablecaption{Summary of all SN models used in this work.} 
\label{tab:AllModels}
\tablehead{
\colhead{Name in} & \colhead{Brief description and publication of origin} & \colhead{Name in} & \colhead{$\mathrm{M_{ej}}^{a}$} \\[-2ex]
\colhead{this work} & \colhead{} & \colhead{original work} & \colhead{(in $M_{\odot}$)}
}
\startdata
DDT   & Deflagration to detonation transition, with one point of ignition \citep{Seitenzahl+2013MNRAS} &  N1        & 1.44 \\
GCD   & Gravitationally confined detonation \citep{Lach+2022AandA} &  r51\_d4.0 & 1.44     \\
DEF   & Deflagration with five points of ignition \citep{Fink+2014MNRAS} &  N5def     & 0.37  \\
DDCO  & Double detonation with 1.0\(M_\odot\)\ core and 0.05\(M_\odot\)\ He shell \citep{Gronow+2021AandA} &  M10\_03   & 1.03  \\
DDONe & Double detonation of ONe WD \citep{Marquardt+2015AandA} &  ONe10e7   & 1.18     \\
VM    & Violent merger of $1.1M_{\odot}$ and $0.9M_{\odot}$ WDs \citep{Pakmor+2012ApJ} &  -$^{b}$   & 1.97 \\
\enddata
\tablenotetext{a}{mass of the ejecta; excludes mass of the bound remnant wherever present}
\tablenotetext{b}{only model in the paper.}
\end{deluxetable*}

\subsection{Modeling the remnant phase} \label{subsec:remnant}

We map the 1D SN models onto an octant of a sphere, with a fiducial domain resolution of $512$ zones per dimension. These were then allowed to expand against a uniform AM on \sprout, a second-order expanding mesh hydrodynamics code \citep{Mandal+2023_sprout}. \sprout\, employs the moving mesh methodology \citep{Springel2010MNRAS,Duffell+2011ApJS} to expand its Cartesian mesh with time, tracking the expansion of the remnant. Here, we solve the equations of ideal, non-relativistic hydrodynamics in three dimensions:

\vspace{-4mm}

\begin{equation}
\label{eq:euler}
    \begin{gathered}
        \partial_t(\rho) + \nabla \cdot ( \rho \mathbf{v} ) = 0 \\
        \partial_t( \rho \mathbf{v} ) + \nabla \cdot ( \rho \mathbf{vv} + P \overleftrightarrow{I} ) = 0 \\
        \partial_t\left( \frac{1}{2}\rho v^2 + \epsilon \right) + \nabla \cdot \left( \left( \frac{1}{2}\rho v^2 + \epsilon + P \right)\mathbf{v} \right) = 0 \\
        \partial_t(\rho X) + \nabla \cdot ( \rho X \mathbf{v} ) = 0,
    \end{gathered}
\end{equation}

where $\rho$, $\mathbf{v}$, $P$, $\epsilon$ are the density, velocity, pressure, internal energy of the fluid respectively, and $X$ is a passive scalar that advects with the fluid. $P$ is related to $\epsilon$ via an adiabatic equation of state: $P = (\gamma-1)\epsilon$, where $\gamma=5/3$ is the adiabatic index. In this problem, we have used four passive scalars, with the first three corresponding to the mass fraction of Si, S and Fe. The fourth passive scalar is used to track the ionization age of the shocked plasma, which is defined as:

\begin{equation}
\label{eq:ionization_age}
   \tau (t) \equiv \int_{t_{\mathrm{shock}}}^t n_e dt, 
\end{equation}

with $n_e$ being the electron density and $t_{\mathrm{shock}}$ is the time since the plasma was shocked. We found that the plasma temperature always exceeds $10^6\,\mathrm{K}$ once it has been shocked \citep[similar to][]{Warren+2013MNRAS}. This was used to keep track of shocked elements, as well as the corresponding $t_{\mathrm{shock}}$ value. The electron density $n_e$ is computed using a technique similar to that of \cite{Dwarkadas+2010MNRAS} (see their equations~4-6). This calculation requires the abundance, and the average number of free electrons for each element. The former is available from the SN models. Here, we approximate the latter using ionization equilibrium calculations \citep{Bryans+2006ApJS,Bryans+2009ApJ} for a fixed plasma temperature of $10^7\,\mathrm{K}$. The ionization age is important for estimating the X-ray emission from our hydrodynamical models, as will be elaborated in section \ref{subsec:synthetic_im}.

The dynamical properties and substructure morphology of our models can be expressed as functions of a single dimensionless quantity called the dynamical age (as opposed to exploring a multi-dimensional parameter space of explosion energy, ejecta mass, ambient density, and age of the remnant). We define the dynamical age of a remnant as its true age divided by the characteristic time or Sedov time ($T_c$), which is the time required for the forward shock to sweep up mass comparable to the ejecta mass itself. The characteristic time is found to be a function of the ejecta mass $M$, explosion energy $E$, and AM density $\rho$ \citep{Warren+2013MNRAS,Mandal+2024ApJ}. We adopt the definition for $T_c$ as in \cite{Truelove+1999ApJS} and \cite{Mandal+2024ApJ}, leading to the following expression for the dynamical age $t_{D}$:

\begin{equation}
\label{eq:dynamical_age}
        t_{D} \equiv \left(\frac{t}{563\mathrm{yrs}}\right)\left(\frac{M}{M_{ch}}\right)^{5/6} E_{51}^{-1/2}n_0^{-1/3},
\end{equation}

where $M_{ch}=1.4M_{\odot}$ is the Chandrasekhar mass, $E_{51}=E/(10^{51}\mathrm{ergs}$), and $n_0 =\rho/(2.34\times10^{-24}\mathrm{g}$) is the number density of the circumstellar medium, assuming a 10:1 H:He ratio. Note that this definition is not unique. For example, \cite{Warren+2013MNRAS} choose a different characteristic age $T'$, which is related to our choice as $T' = 0.43T_c$. Our models are evolved up to a dynamical age of 1.0. This is  likely well beyond that of Tycho's SNR, which is expected to have a dynamical age in the range $0.4-0.8$. \cite{Mandal+2024ApJ} deduce $t_D=0.5$ for Tycho's SNR from analysis of its small-scale substructures. A similar study by \cite{Warren+2013MNRAS} (which also includes kinematic comparisons) suggest a dynamical age of $0.4-0.5$ for both the Tycho and SN 1006 remnants (taking into account the difference in definition of the characteristic age compared to ours; see above). \cite{Badenes+2006ApJ} find an explosion energy of $1.2\times10^{51}\mathrm{erg}$, ejecta mass of $1.4M_{\odot}$, and an ambient density of $2\times10^{-24}\mathrm{gcm^{-3}}$ for Tycho's SNR, from their 1D hydrodynamic simulations and detailed X-ray flux calculations. These values yield $t_D\approx0.74$.

\subsection{Model analysis} \label{subsec:model_anly}

Power spectral analysis of our 3D models is performed as in \cite{Polin+2022ApJ} and \cite{Mandal+2023ApJ}. Power spectra are obtained for all our models at several instants of time. Given a model snapshot, we calculate spherical surface distributions of the radially integrated ejecta density and density distribution of a given element M as follows:

\begin{equation}
    \begin{split}
        \left<\rho_{\mathrm{ej}}\right>(\theta,\phi) = \frac{\int P \rho dr}{\int P dr} \,, \\
        \left<\rho_M\right>(\theta,\phi) = \frac{\int X_M P \rho dr}{\int X_M P dr}, \\
    \end{split}
\label{eq:surface_maps}
\end{equation}

where $X_M$ is the passive scalar corresponding to the mass fraction of element M (M being either Si, S or Fe). Weighting by pressure ensures only fluid from the shocked, turbulent region is considered. These distributions are then normalized by dividing with their angle-averaged value. The relative amplitude of anisotropies in, say the density distribution of Fe, may be obtained by expanding it as a sum of spherical harmonics:

\begin{equation}
    \left<\rho_{\mathrm{Fe}}\right>(\theta,\phi) = \sum\limits_{l,m} a_{lm} Y_{lm}(\theta,\phi) \,.
\end{equation}

The amplitudes of the harmonics are used to calculate a power spectrum as a function of the harmonic number $l$:

\begin{equation}
    C_l = \frac{1}{2l+1}\sum_{m=-l}^{+l}\left|a_{lm}\right|^2 \,.
\end{equation}

We obtain power spectra for our normalized surface maps using the SHTOOLS package \citep{SHTOOLS}. The power spectra presented in this work are a smoothened version of the raw power spectra. These were obtained by convolving the raw versions with a Gaussian kernel as in \cite{Polin+2022ApJ}. All analysis was performed on the raw versions.

\subsection{Synthetic images} \label{subsec:synthetic_im}

Computing synthetic X-ray images from hydrodynamical models is necessary for comparison against high resolution X-ray images of SNRs. X-ray emission in SNRs originates primarily in shock-heated plasma and includes both thermal and non-thermal components. The thermal component consists of continuum emission and line emission; the reader is referred to \cite{Vink2012A&ARv} for a general review. In this work we focus on the line emission produced by collisional excitation of ions. Electron–ion collisions excite ions, which subsequently decay radiatively to lower energy states. As mentioned in section~\ref{subsec:initial}, our elements of interest are Si, S, and Fe. These elements produce line emission in certain narrow energy bands, which will be discussed in Section~\ref{sec:observations}. Images in these narrow bands thus primarily trace emission from the corresponding collisionally excited ions.


The line emissivity depends on the ionic charge state distribution, which is governed by ionization and recombination in the shocked plasma. For the typical densities in SNRs ($\lesssim10^5\,\mathrm{cm^{-3}}$), young remnants do not reach collisional ionization equilibrium (CIE) and instead remain in a non-equilibrium ionization (NEI) state \citep{Itoh1977PASJ,Hamilton+1983ApJS,Borkowski+2001ApJ}. As a rule of thumb, CIE is approached only for ionization ages $\tau\gtrsim10^{12}\,\mathrm{cm^{-3}\,s}$ \citep{Vink2006ESASP,Dwarkadas+2010MNRAS}. Recently shocked gas therefore emits weakly in line emission despite often having high density.


In this work, we construct approximate X-ray narrowband images of SNRs intended to trace the spatial distribution of emitting ejecta, rather than performing detailed NEI spectral calculations. We impose a simple visibility criterion based on ionization age: zones with $\tau\le10^{12}\,\mathrm{cm^{-3}\,s}$ are treated as X-ray faint (zero emissivity), while zones with larger $\tau$ are assigned emissivity proportional to the square of the density of the selected element (computed from the gas density and elemental abundance). Synthetic images are then obtained by integrating this emissivity along parallel lines of sight.

This prescription is not intended to reproduce the detailed plasma conditions of a particular remnant such as Tycho's SNR, whose ejecta have ionization ages in the range $10^{10}\mbox{--}10^{11}\,\mathrm{cm^{-3}\,s}$ \citep{Yamaguchi+2014ApJ,Lu+2015ApJ}. Instead, the aim is to suppress emission from freshly shocked material near the forward shock that would otherwise dominate a simple density-squared projection. The resulting maps are therefore interpreted as tracers of observed ejecta morphology rather than absolute surface brightness. The precise threshold is not critical: varying it within $10^{10}\mbox{--}10^{12}\,\mathrm{cm^{-3}\,s}$ produces nearly identical morphologies, since it primarily excludes recently shocked gas rather than selecting a specific ionization state. Examples of synthetic images (from the DDCO model) corresponding to Si, S, and Fe emission are shown in the lower panels of Fig.~\ref{fig:xray_images}, with images of Tycho’s SNR shown in the upper panels for qualitative comparison.

\subsection{Image analysis} \label{subsec:image_anly}

We perform substructure analysis of the \textit{Chandra} X-ray images and our synthetic images (as in Fig.~\ref{fig:xray_images}). We employ the $\Delta$-variance algorithm \citep{Arevalo+2012MNRAS}, as adapted by \cite{Mandal+2024ApJ}. This algorithm computes a low-resolution power spectrum of a 2D image, with power evaluated as a function of spatial scale $\sigma$ (in pixel units).

To enable a direct comparison between models and observations, all images (both synthetic maps and observations) are analyzed in an identical manner. Each image is centered on the remnant. The remnant radius $R_{\mathrm{SNR}}$ is determined separately for each image from the outer boundary of detectable emission. The power spectrum is then expressed in terms of the angular harmonic $l$ (in the form described in section~\ref{subsec:model_anly}) using the following relation between $l$ and $\sigma$ \citep{Mandal+2024ApJ}:

\vspace{-2mm}

\begin{equation}
\label{eq:sigma_to_l}
    l\approx 1.6R_{\mathrm{SNR}}/\sigma,
\end{equation}

where both $R_{\mathrm{SNR}}$ and $\sigma$ are expressed in pixel units. Because the analysis is performed in terms of the dimensionless harmonic $l$, the results are independent of the absolute pixel scale or physical size of the remnant. No additional rescaling between simulations and observations is required. The comparison is therefore strictly morphological and scale-normalized.

\begin{figure*}
\centering
\includegraphics[width=0.98\textwidth]{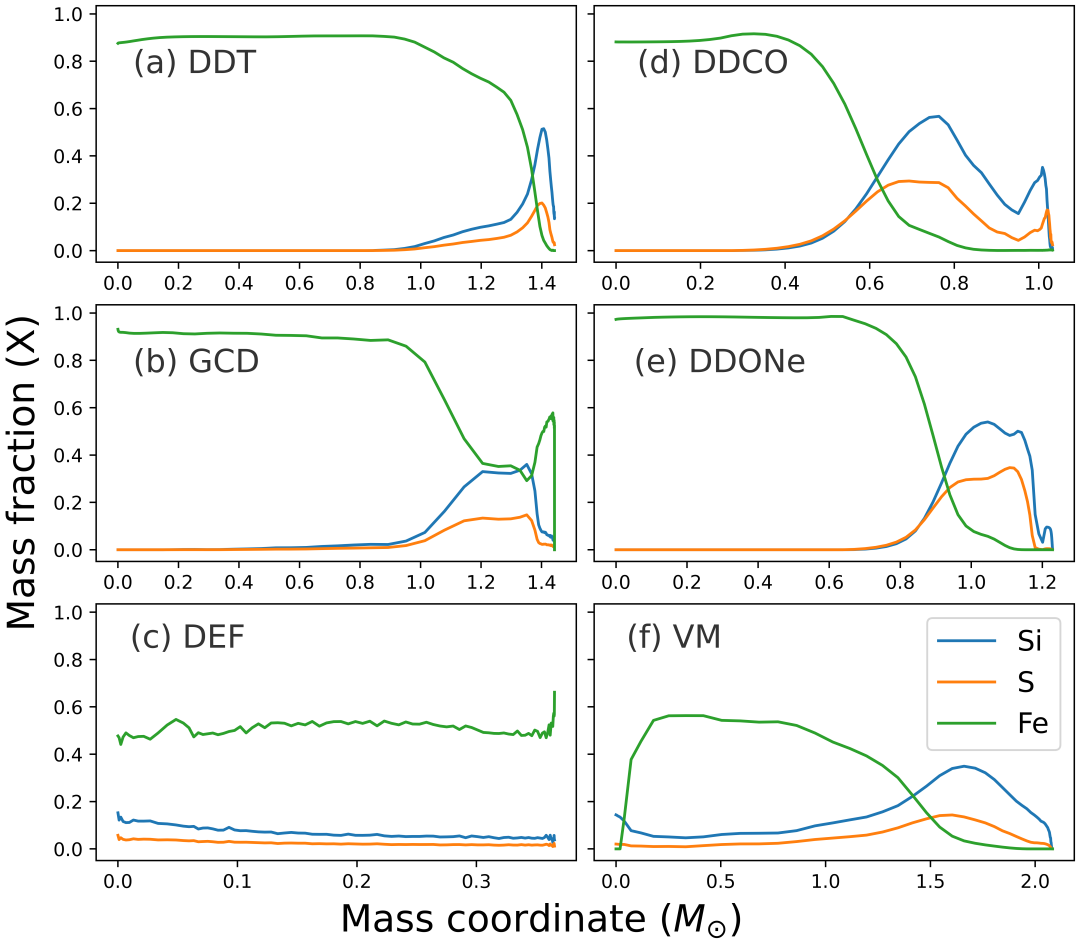}
\vspace{-1mm}
\caption{Si, S, and Fe mass fraction as a function of mass coordinates for all SN models that have been used as initial conditions for this work. DDT, GCD, and DEF are near-$\mathrm{M_{ch}}$ models, while the remaining are sub-$\mathrm{M_{ch}}$ models. Further details may be found in Table~\ref{tab:AllModels}. }
\label{fig:model_abundance_profiles}
\end{figure*}

\section{Observational data reduction}
\label{sec:observations}

There are 23 \textit{Chandra} ACIS-I/S observations without grating of Tycho's SNR in the archive, which amount to a total of 1.4~Ms of exposure time. However, comparison between observations of the remnant at different times (the observations span from the year 2000 to the year 2022) have shown that it expands in real time. For this reason, stacking all of the existent data for the remnant could result in blurring out the smaller structures, artificially creating larger clumps. 

In order to prevent this effect, we opt to stack only observations taken in 2009. The reason for this choice is that all 2009 observations were taken in the span of less than 1 month, yet they amount to 0.74~Ms of exposure time, roughly half of the total available. All observations used in this work are contained in the \textit{Chandra} Data Collection (CDC) 338~(\href{https://doi.org/10.25574/cdc.338}{doi:10.25574/cdc.338}), and their characteristics can be found in Table \ref{tab:AllObservations}. 

\begin{deluxetable}{ccccccc}
\tablecaption{Summary of all \textit{Chandra} observations analyzed in this work.} 
\label{tab:AllObservations}
\tablehead{\colhead{Date} & \colhead{Obs ID} & \colhead{Exp. Time (ks)}   \\ \colhead{(1)} & \colhead{(2)} & \colhead{(3)}  }
\startdata
2009-04-11 & 10097 & 108.9  \\
2009-04-13 & 10904 & 35.2   \\
2009-04-13 & 10093 & 119.9 \\
2009-04-15 & 10902 & 40.1  \\
2009-04-17 & 10903 & 24.2 \\
2009-04-18 & 10094 & 91.2 \\
2009-04-23 & 10095 & 175.7  \\
2009-04-27 & 10096 & 107.1 \\
2009-05-03 & 10906 & 41.7 \\
\enddata
\tablenotetext{}{\textbf{Notes:} (1) Observation date. (2) Observation ID. (3) Exposure time.}
\end{deluxetable}

\begin{figure}
\centering
\includegraphics[width=0.48\textwidth]{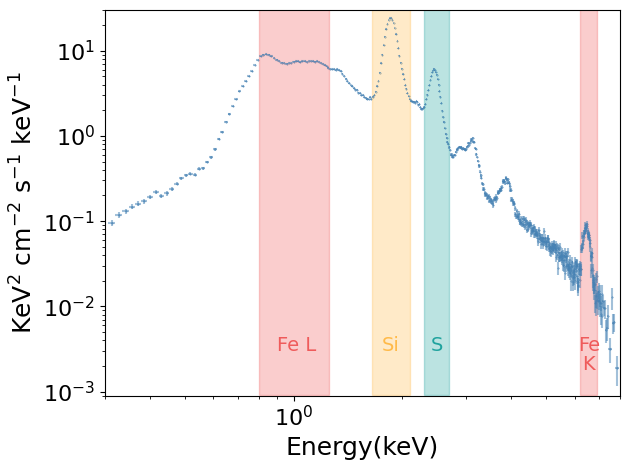}
\caption{\textit{Chandra} spectrum taken from Obs ID 10093, illustrating the four different energy bands used in this work, and how they correspond to different line emission: Fe L in 0.8-1.25 keV (pink), Si in $1.65-2.1$ (yellow), S in $2.3-2.7$ (green), and Fe K$_{\alpha}$ in $6.2-6.9$ (pink) keV.}
\label{fig:chandra_spec}
\end{figure}

We merged the 9 available observations following the corresponding CIAO thread\footnote{https://cxc.cfa.harvard.edu/ciao/threads/combine/}, and the CIAO tool \texttt{merge\_obs}. 
We then created images at different energy bands corresponding to the distribution of different elements using the CIAO tool \texttt{dmcopy}. As Fig.~\ref{fig:chandra_spec} shows, the bands $1.65-2.1$ keV and $2.3-2.7$ keV correspond to Si and S emissivity distributions, respectively \citep{Badenes+2006ApJ}. For iron, we consider both the Fe L ($0.8-1.25$ keV) and the Fe K$\alpha$ ($6.2-6.9$ keV) bands. The Fe K$\alpha$ band provides a reliable tracer of Fe ejecta in young Type Ia SNRs because line emission in this band originates K-shell transitions in highly ionized Fe heated by the reverse shock and is minimally contaminated by other species. However, the photon statistics in this band are comparatively limited. The Fe L band, by contrast, provides substantially higher photon statistics. Although this band consists of a blend of multiple L-shell transitions, contamination from IMEs in this band is expected to be minor for young Type-Ia SNRs, as demonstrated in Fig.~10 of \cite{Yamaguchi+2017ApJ}. We therefore extract images in both Fe L and Fe K$\alpha$ bands. The Fe L images ensure that small-scale structures are traced with high statistical fidelity, mitigating potential bias in the power-spectrum slope due to counting noise. The Fe K$\alpha$ images, while photon-poor, provide an independent and physically robust tracer of Fe-rich ejecta. Merged and exposure-corrected \textit{Chandra} images are displayed in Fig.~\ref{fig:xray_images} for the Si, S and Fe bands indicated above. 



\section{Results}   \label{sec:results}

\subsection{Features of the density power spectrum}
\label{subsec:model_ps}

As mentioned in section~\ref{subsec:model_anly}, each model snapshot provides us with four angular power spectra, corresponding to the ejecta density, and density distributions of Si, S and Fe. All of these are found to exhibit the characteristic broken power-law shape as found by previous studies \citep{Polin+2022ApJ,Mandal+2024ApJ}. In this section, we discuss features of the density power spectrum (agnostic to the composition of the ejecta). As an example, time evolution of the density power spectrum for the DDCO model is shown in the top panel of Fig.~\ref{fig:density_ps_dd}. The characteristic wavenumber where the (power-law) slope of the density power spectrum changes value is henceforth referred to as $l_0$, following previous works. In our models, this is also where the power spectrum peaks, given the absence of strong large-scale asymmetries. Therefore $l_0$ represents the length scale where most of the turbulent kinetic energy in the SNR resides. We fit density power spectrum from all our model snapshots to the following broken power-law expression \citep{Mandal+2024ApJ}:

\begin{figure}
\centering

\includegraphics[width=0.48\textwidth]{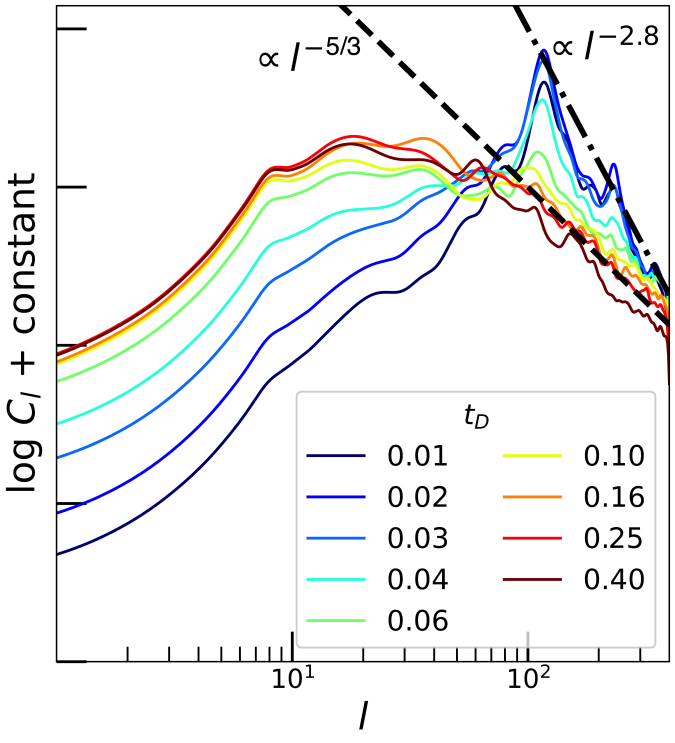}
\includegraphics[width=0.50\textwidth]{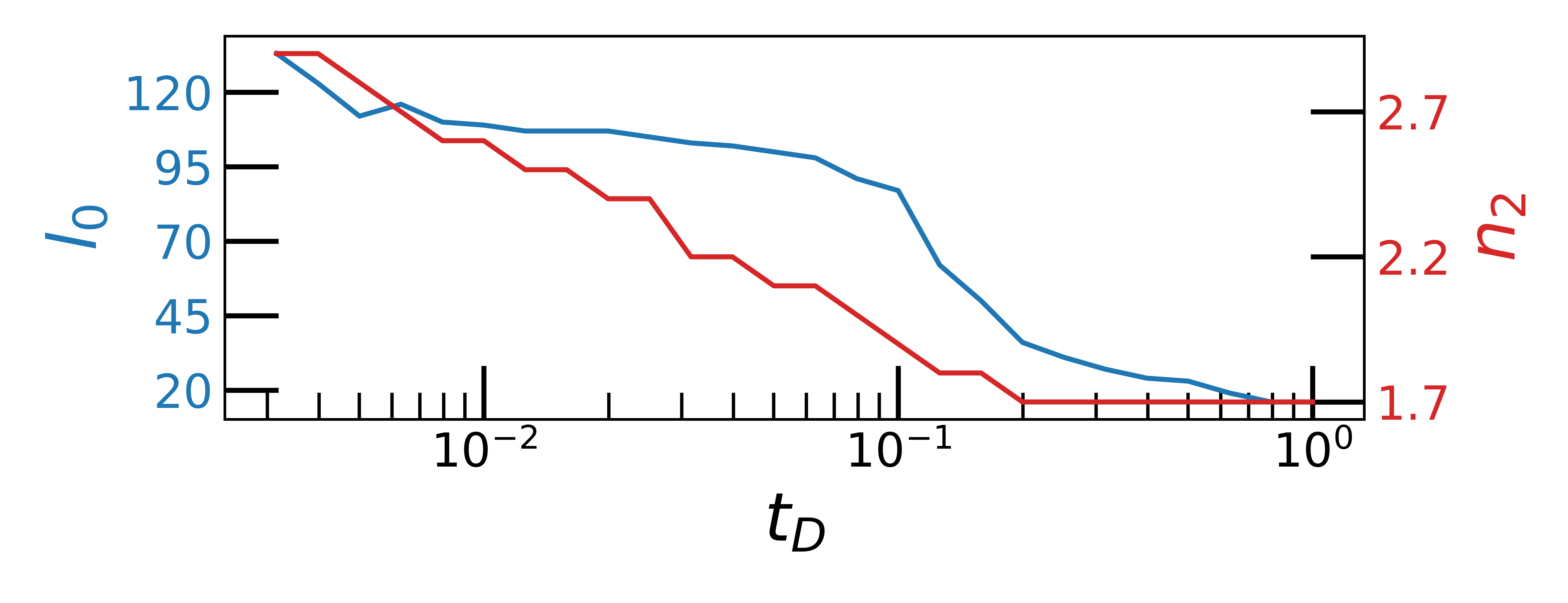}
\caption{\textbf{Top:} Density power spectra of the DDCO model evaluated at 9 logarithmically spaced time instants between $t_D=0.01-0.4$ or $t\approx5-200\mathrm{yr}$, assuming a typical Sedov time of 500 years (see Eqn.~\ref{eq:dynamical_age}). All power spectra have a broken power-law shape. The wavenumber where the power spectrum breaks ($l_0$) corresponds to the characteristic physical scale of the turbulent substructures. With time, $l_0$  shifts to smaller values, indicating growth of RTI structures. The power spectrum is initially steep for large values of the wavenumber $l$ ($C_l\propto l^{-2.8}$) but transitions to a shallower shape ($C_l\propto l^{-1.7}$) at late times, which is consistent with the Kolmogorov energy cascade. \textbf{Bottom:} The peak wavenumber $l_0$ and the large-$l$ power-law slope $n_2$ (see Eqn.~\ref{eq:fit}) as a function of $t_D$. }
\label{fig:density_ps_dd}
\end{figure}

\begin{equation}
\label{eq:fit}
    P_l = \frac{A}{(l/l_0)^{-n_1}+(l/l_0)^{n_2}}.
\end{equation}

The fit function in Eq.~\ref{eq:fit} implies that the power spectrum can be described approximately as $C_l \propto l^{n_1}$ and $C_l \propto l^{-n_2}$ for $l<l_0$ and $l>l_0$, respectively. Therefore, the fit provides the dominant angular mode or the break wavenumber $l_0$ and the power-law slopes $n_1$ and $n_2$ for each model snapshot. The right panel of Fig.~\ref{fig:density_ps_dd} plots $l_0$ and $n_2$ for the DDCO model as a function of the dynamical age, $t_D$. It is seen that $l_0$ asymptotes to $\sim15$ by $t_D=1.0$. We return to the behavior of $n_2$ later in this section.

Previous studies \citep{Polin+2022ApJ,Mandal+2023ApJ,Mandal+2024ApJ} have demonstrated that $l_0$ is directly proportional to the power-law slope of the outermost ejecta layers (that is, ejecta that the reverse shock interacts with at the time of investigation). For ejecta with a power-law density profile ($\rho \propto r^{-n}$), the density gradient in the radial direction is:

\vspace{-4mm}

\begin{equation}
\label{eq:rho_gradient}
        \frac{d\rho}{dr} = - \frac{n \rho(r)}{r}\,, \\
\end{equation}

with $n$ being the power-law index or slope. We use Eq.~\ref{eq:rho_gradient} to define an effective power-law slope, $n_{\mathrm{eff}}$ at the position of the reverse shock (RS) as follows:

\begin{equation}
\label{eq:n_eff_def}
    n_{\mathrm{eff}} \equiv \left| \frac{r}{\rho}\frac{d \rho}{d r}\right|_{r=r_{\mathrm{RS}}} \,.
\end{equation}

\begin{figure}
\centering
\includegraphics[width=0.45\textwidth]{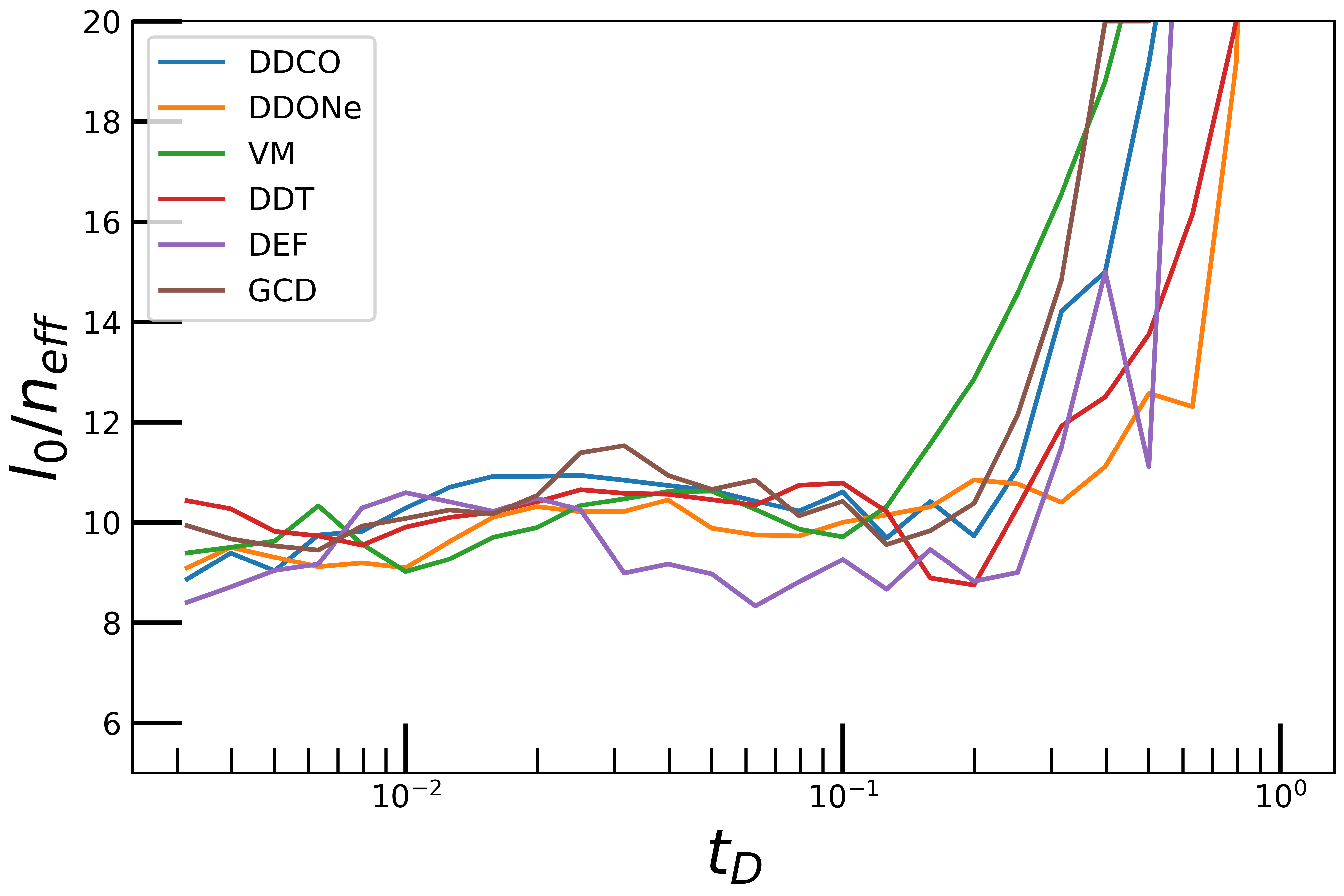}
\caption{Ratio of the peak wavenumber ($l_0$) to the effective power-law slope ($n_{\mathrm{eff}}$) of the outermost ejecta encountered by the reverse shock, plotted against time for all SNR models. All models are found to be consistent with the relation $l_0/n_{\mathrm{eff}} \sim 10$ as found by \cite{Mandal+2024ApJ} except at very late times.}
\label{fig:evolution}
\end{figure}

In Fig.~\ref{fig:evolution}, we plot the time evolution of the quantity $l_0/n_{\mathrm{eff}}$ for all our SNR models. This ratio is seen to be roughly constant ($l_0/n_{\mathrm{eff}} \sim 10$) for all models except at very late times. \cite{Mandal+2024ApJ} find essentially the same result for their exponential density profile ejecta. In both cases, $l_0$ asymptotes to a constant at late times while $n_{\mathrm{eff}}$ keeps decreasing, thus blowing up $l_0/n_{\mathrm{eff}}$. The result $l_0 \propto n$ implies that the characteristic size of the RTI substructures is of the order of the density scale height of the ejecta, as shown by \cite{Polin+2022ApJ}. 


The other feature of interest in the power spectrum is its shape for $l>l_0$ (corresponding to length scales smaller than that for $l_0$). According to the classical theory of turbulence \citep{Kolmogorov1941DoSSR}, an energy cascade from larger to smaller length scales is expected, which would imply a power-law, $C_l \propto l^{-5/3}$. Previous studies \citep{Polin+2022ApJ,Mandal+2023ApJ} employing SN ejecta with a power-law density profile found a much steeper power-law ($C_l \propto l^{-3}$) for $l>l_0$, which is speculated to indicate suppression of turbulent cascade due to the expansion of the blastwave. We find that the density power spectrum for our models start with a steep shape and gradually transition to the Kolmogorov cascade shape ($C_l \propto l^{-5/3}$) at late times, as shown for the DDCO model in Fig.~\ref{fig:density_ps_dd} (top panel). This is depicted more clearly in the bottom panel, where we see that the large-scale power-law slope $n_2$ asymptotes to $\approx1.7$ by $t_D\approx0.1$. This was also noted by \cite{Mandal+2024ApJ}, again depicting the success of the W7 model (or exponential density profile) in reproducing general features of the turbulence in other Type-Ia SN models.

\begin{figure*}
\centering
\includegraphics[width=0.98\textwidth]{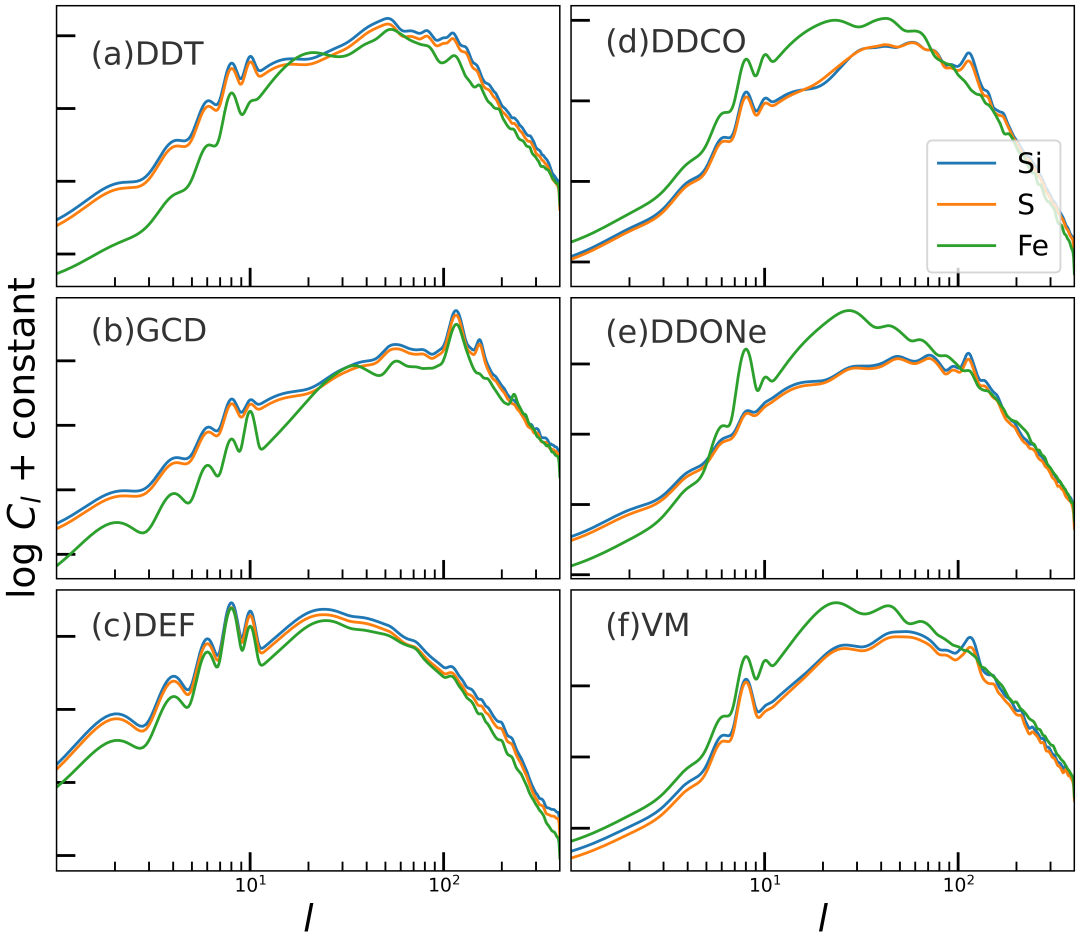}
\vspace{-3mm}
\caption{Power spectra of Si, S, and Fe distribution for all SNR models at $t_D=0.5$. The near-$\mathrm{M_{ch}}$ (DDT, DEF, and GCD) models have power spectra that peak at almost the same wavenumber. In contrast, the power spectra of Fe distribution in sub-$\mathrm{M_{ch}}$ (DDCO, DDONe, and VM) models peaks at a smaller wavenumber compared to the power spectra of Si or S distributions. Therefore typical Fe-dominated substructures are expected to be larger than their Si or S counterparts in SNRs that originated from a sub-$\mathrm{M_{ch}}$ WD.}
\label{fig:ps_of_elements}
\end{figure*}

\subsection{Power spectra of Si, S, and Fe distributions}
\label{subsec:PS_of_elements}

Fig.~\ref{fig:ps_of_elements} shows power spectra of density distributions of Si (blue), S (orange) and Fe (green) for all our SNR models, at a dynamical age of $t_{D}=0.5$, which falls in the likely range of dynamical age for Tycho's SNR (see section~\ref{subsec:remnant}). The left and the right panels show results for the near-$\mathrm{M_{ch}}$ (DDT, DEF, and GCD) models and the sub-$\mathrm{M_{ch}}$ (DDCO, DDONe, and VM) models, respectively. All of these power spectra reflect the characteristic broken power-law shape, but it is immediately apparent that the power spectra for Si, S and Fe do not peak at the same angular mode for all models. It's also clear that the peak angular modes of different elements (referred to as $l_0^{\mathrm{Si}}$, $l_0^{\mathrm{S}}$ and $l_0^{\mathrm{Fe}}$ henceforth) are very similar for the near-$\mathrm{M_{ch}}$ models, while sub-$\mathrm{M_{ch}}$ models have $l_0^{\mathrm{Fe}}>l_0^{\mathrm{S}}\approx l_0^{\mathrm{Si}}$. In other words, Fig.~\ref{fig:ps_of_elements} shows us that the typical size of Si-dominated or S-dominated substructures can be different from that of Fe-dominated substructures, which is what is seen in Fig.~1 of \cite{Lu+2011ApJ}. 

However, this is not trivial to conclude: the power spectra presented up to this point are derived from 3D data, while substructures observed in SNRs are from 2D images, which suffer from line-of-sight projection effects. \cite{Mandal+2024ApJ} show that angular power spectra obtained from 3D SNR models are equivalent (except at very large scales) to the power spectra derived using $\Delta$-variance from the synthetic 2D images of their 3D models (see their Figure 6). We verify the same for our work by calculating the power spectrum of synthetic narrowband images from our models. In section \ref{subsec:tycho_ps}, we present power spectra obtained from narrowband images of Tycho's SNR, corresponding to Si, S and Fe emission. These are compared to the best-match power spectra obtained from the synthetic images of our models.

\begin{figure*}
\centering
\gridline{\fig{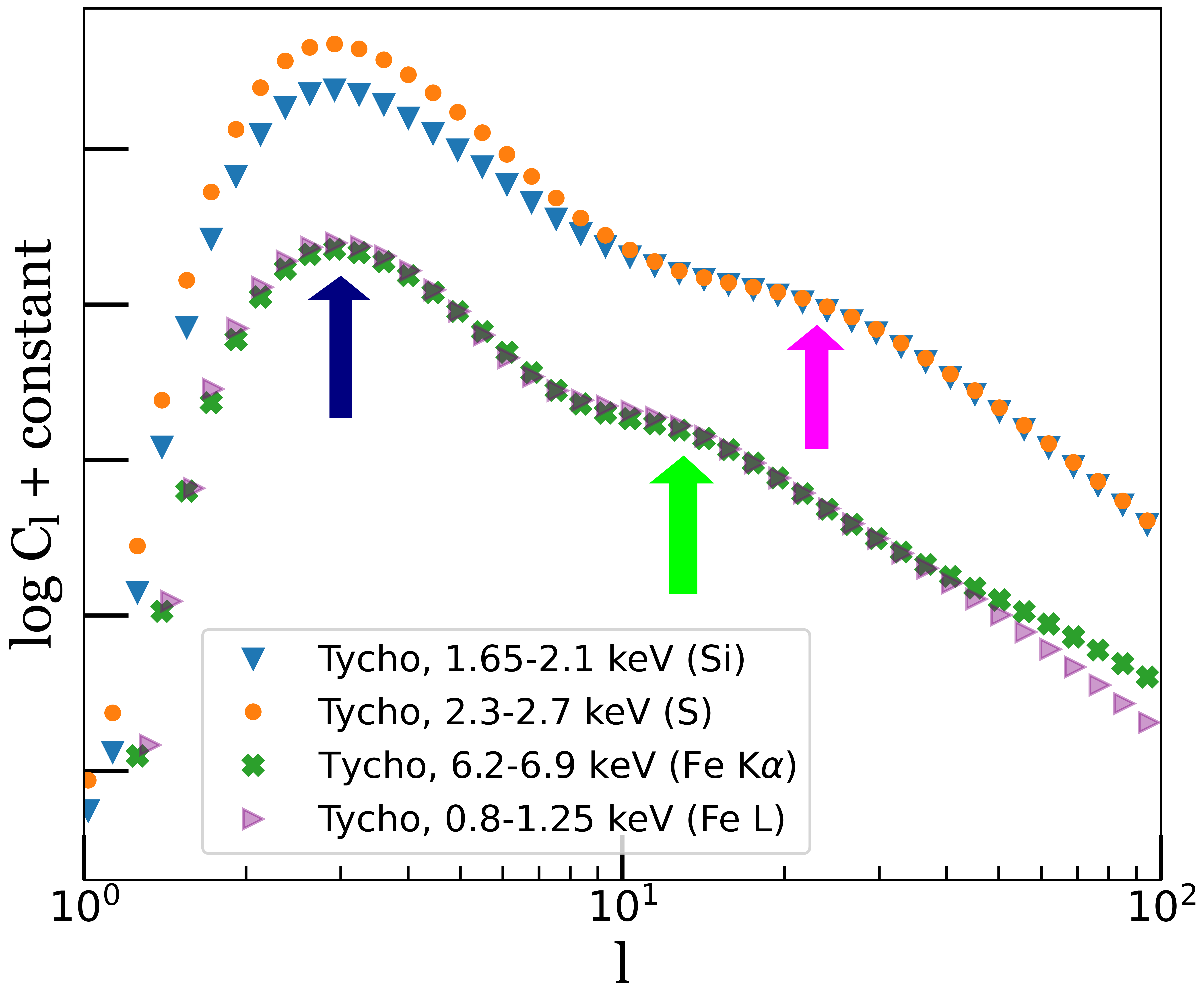}{0.49\textwidth}{}\fig{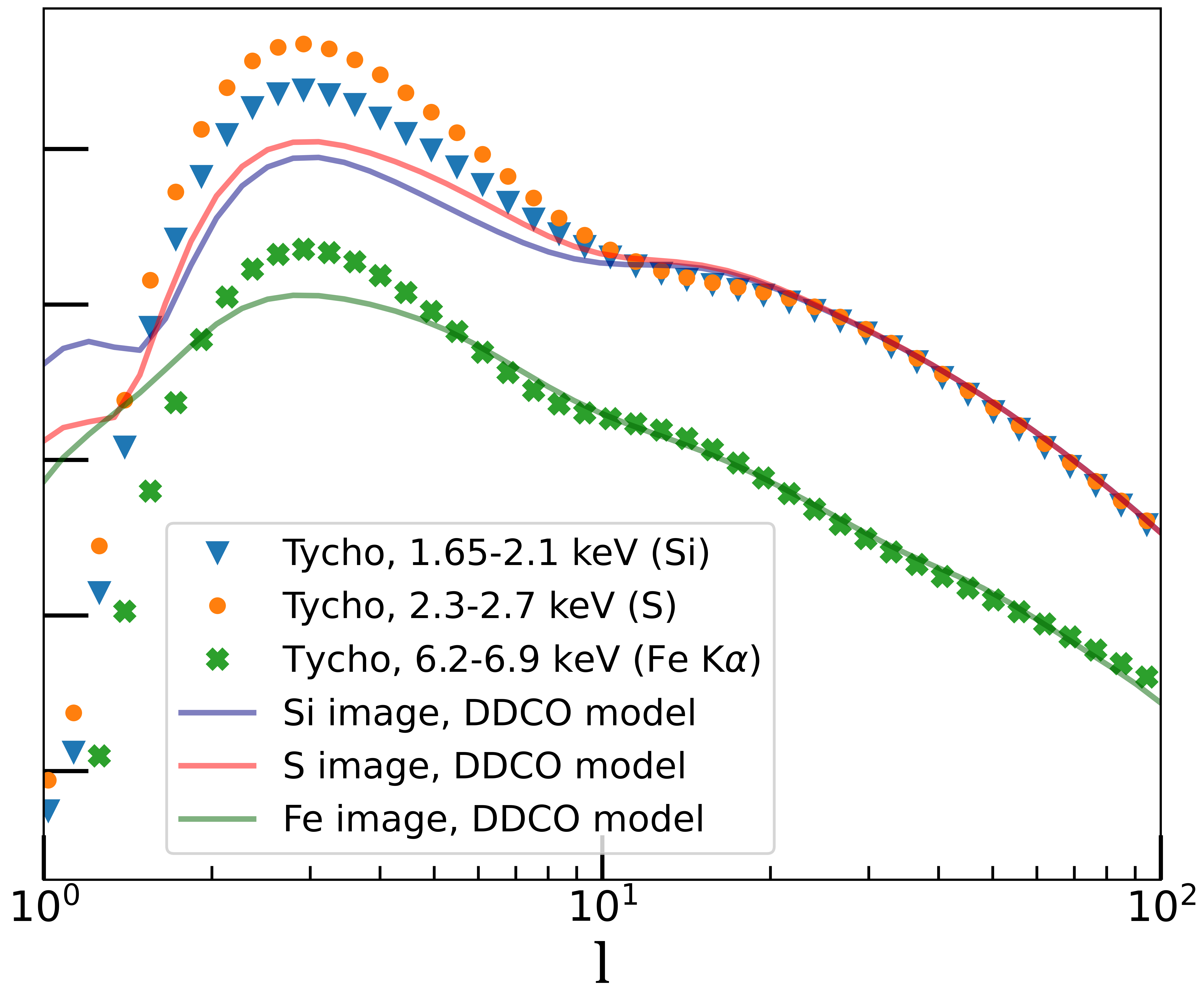}{0.49\textwidth}{}}
\vspace{-7mm}
\caption{\textbf{Left:} Power spectra of Si, S, Fe K$\alpha$, and Fe L emissivity distributions in Tycho's SNR (bottom-pointing triangle, dot, cross, and right-pointing triangle markers, respectively). The power spectra for Si and S show a break in shape at $l\sim23$ (marked with a magenta arrow), while the power spectrum for Fe shows a similar break at $l\sim13$ (marked with a green arrow). As in \cite{Mandal+2024ApJ}, we identify this break with the break wavenumber due to RTI activity. This behavior is consistent with the sub-$\mathrm{M_{ch}}$ models but not with the near-$\mathrm{M_{ch}}$ models (see Figs.~\ref{fig:ps_of_elements} and \ref{fig:l0_ratio_ev}). There's also a peak at a smaller wavenumber ($l\sim3$; marked with a navy arrow), which is attributed to projection effects and a possible large scale anisotropy inherent to the SN ejecta. \textbf{Right:} These power spectra are best matched by those of synthetic Si, S, and Fe X-ray images of the DDCO model, at $t_D=0.6$. The power spectrum of Tycho in the Fe L band is left out due to its similarity to the same for Fe K$\alpha$. A more detailed comparison against other models and other time instants is presented in Figs.~\ref{fig:models_at_tD_0p6} and~\ref{fig:tycho_vs_ddco}. Note that the power spectra of synthetic images are not necessarily the same as that from the 3D models (as in Figs.~\ref{fig:density_ps_dd} or~\ref{fig:ps_of_elements}) due to projection effects. The power spectra were normalized arbitrarily to facilitate visual inspection.}
\label{fig:tycho_ps}
\end{figure*}

\subsection{Power spectra of Tycho's SNR}
\label{subsec:tycho_ps}

Fig.~\ref{fig:tycho_ps} shows the power spectra of the Si, S and Fe emissivity distributions in Tycho. We note that the overall normalizations of these power spectra are arbitrary. They have been chosen to avoid crowding in the plots and thereby facilitate visual inspection. The power spectra are similar to those  obtained by \cite{Mandal+2024ApJ} for a broadband \textit{Chandra} X-ray image of Tycho's SNR. The characteristic strong peak at low wavenumbers ($l\approx3$) hints at projection effects and a possible large-scale asymmetry (such as the bright arc in the northwestern part of Tycho). This feature is therefore absent from the angular power spectra derived from our 3D models, as in Figs.~\ref{fig:density_ps_dd} or \ref{fig:ps_of_elements}. Following \cite{Mandal+2024ApJ}, we interpret the bend in the power spectra for Si and S (corresponding to the $1.6\mbox{--}2.1$ keV and the $2.3\mbox{--}2.7$ keV images) around $l\approx23$ as the break wavenumber $l_0$ for the power spectrum of Si and S. A similar size of substructures was found by \cite{Lopez+2011ApJ}, also for a broadband image of Tycho's SNR \citep[se comparison in][]{Mandal+2024ApJ}. In contrast, the power spectra derived from both the Fe K$\alpha$ ($6.2\mbox{--}6.9$ keV) and the Fe L ($0.8\mbox{--}1.25$ keV) images exhibit a clear change in slope at $l\approx13$. The close agreement between the power spectra obtained from these two independent Fe tracers, including the presence of the same bend at $l\approx13$, indicates that this feature is intrinsic to the spatial distribution of Fe-rich ejecta. In particular, the consistency between the photon-rich Fe L band and the comparatively photon-poor Fe K$\alpha$ band argues against an interpretation in which the small-scale power or the observed bend is driven by statistical noise or contamination from other elements (see discussion in Section~\ref{sec:observations}). We therefore infer $l_0^{\mathrm{Si}}=l_0^{\mathrm{S}}\approx13$ and $l_0^{\mathrm{Fe}}\approx23$ for Tycho. Comparing the power spectra of Tycho to those of our SNR models (Fig.~\ref{fig:ps_of_elements}), we find that only the sub-$\mathrm{M_{ch}}$ models in our sample produce such a significant disparity between $l_0^{\mathrm{Si}}$ (or $l_0^{\mathrm{S}}$) and $l_0^{\mathrm{Fe}}$.

To enable direct comparison between our models and Tycho's SNR, we obtain power spectra of synthetic images derived from each of our SNR models for the range $t_D=0.4-0.8$. This is the likely range of dynamical age of Tycho (see discussion in section~\ref{subsec:remnant}), with Tycho's small-scale substructure analysis favoring $t_D\approx0.5$ \citep{Mandal+2024ApJ}. We emphasize that these power spectra are derived from the synthetic images (using the technique in section~\ref{subsec:image_anly}) and can be different from the power spectra directly derived from models (as in Fig.~\ref{fig:ps_of_elements}). The main difference between model-derived power spectrum and synthetic image-derived power spectrum is that large scale (small $l$) features show up in the latter owing to projection effects \citep[also see][]{Mandal+2024ApJ}. We find that these features can sometimes interfere with the peak at $l_0$, especially at late times. We therefore use only the model-derived power spectra for analysis purposes, leaving the image-derived power spectra for comparisons to data. We find that synthetic images derived from the DDCO model at $t_D=0.6$ provides the best match to Tycho's power spectra, particularly to the values of $l_0^{\mathrm{Si}}$ and $l_0^{\mathrm{Fe}}$ exhibited by Tycho's power spectra. The small scale (large $l$) shapes of Tycho's power spectra are also matched well by those from the DDCO model, which are plotted in Fig.~\ref{fig:tycho_ps} as well. A more detailed comparison against the models is presented in Appendix \ref{app:tycho_vs_models}, where we plot Tycho's power spectra against power spectra from all models at $t_D=0.6$, as well as power spectra from the DDCO model at different time instants. Appendix \ref{app:all_synthetic_images} shows synthetic images from all models at $t_D=0.6$ (except for the DDCO model, which is shown in Fig.~\ref{fig:xray_images}).

Larger structures in Tycho's SNR, particularly those corresponding to the $l\approx3$ mode, are also seen in the power spectra of the DDCO model. The model spectra, however, show enhanced power at the lowest harmonic ($l\approx1$), which is not observed in Tycho. This arises from the imposed mirror symmetry of the numerical setup. The SNR models are computed in a single octant and the full remnant image is reconstructed by reflection across the coordinate planes, introducing a weak large-scale dipole-like mode that appears primarily at $l=1$. This feature reflects the global geometric reconstruction rather than ejecta substructure and does not affect the intermediate-scale modes ($l\gtrsim10$) used in our analysis. The issue of disparity between the typical size of IGE vs IME dominated substructures is discussed further in section~\ref{subsec:l0_ratio}.

\subsection{IGE-to-IME break wavenumber ratio ($l_0^{\mathrm{Si}}/l_0^{\mathrm{Fe}}$)}
\label{subsec:l0_ratio}

\begin{figure}
\centering
\includegraphics[width=0.48\textwidth]{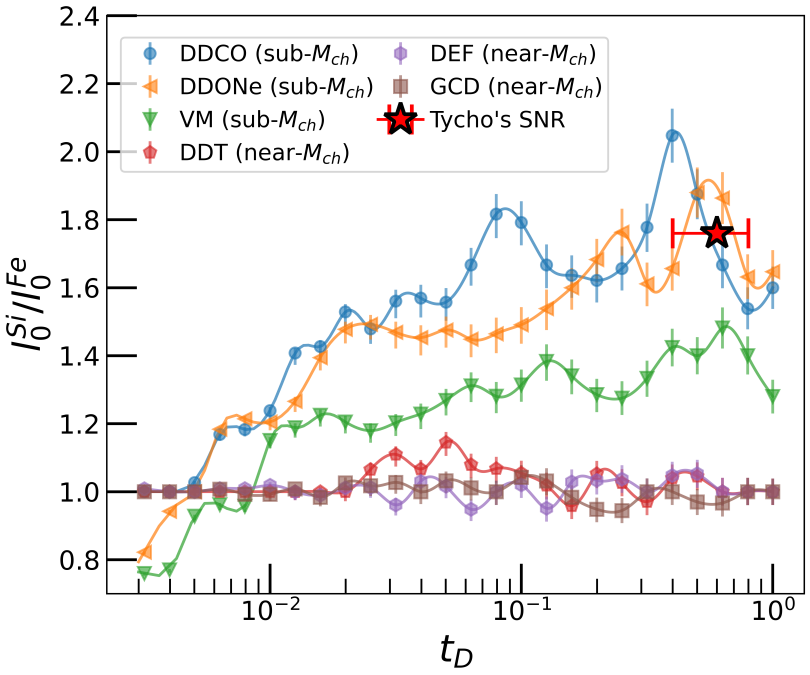}
\vspace{-6mm}
\caption{Time evolution of the ratio $l_0^{\mathrm{Si}}/l_0^{\mathrm{Fe}}$ for all SNR models. $l_0^{\mathrm{Si}}$ and $l_0^{\mathrm{Fe}}$ correspond to the peak wavenumbers of the power spectra of Si and Fe distribution, respectively. Data points are connected by interpolation curves that are only meant to guide the eye. The value of $l_0^{\mathrm{Si}}/l_0^{\mathrm{Fe}}$ measured for Tycho's SNR is overlaid, using estimates of its dynamical age $t_D$. For $t_D\gtrsim 0.1$, the near-$\mathrm{M_{ch}}$ models (DDT, GCD, and DEF) have $l_0^{\mathrm{Si}}/l_0^{\mathrm{Fe}}$ ratios near unity, while the sub-$\mathrm{M_{ch}}$ models exhibit $l_0^{\mathrm{Si}}/l_0^{\mathrm{Fe}}\gtrsim 1.2$.}
\label{fig:l0_ratio_ev}
\end{figure}

In this section, we discuss how the ratio $l_0^{\mathrm{Si}}/l_0^{\mathrm{Fe}}$ (or equivalently $l_0^{\mathrm{S}}/l_0^{\mathrm{Fe}}$) evolves with time for different SNR models. The time evolution of $l_0^{\mathrm{Si}}/l_0^{\mathrm{Fe}}$ is plotted in Fig.~\ref{fig:l0_ratio_ev}. The ratios corresponding to the near-$\mathrm{M_{ch}}$ models hover close to unity at all times, while those for the sub-$\mathrm{M_{ch}}$ models increase beyond unity and consistently remain in the range $1.3-2.0$ at late times. This shows that the typical size of IGE (or Fe) dominated substructures in sub-$\mathrm{M_{ch}}$ SNR models is larger compared to their IME (Si and S here) counterparts, since a smaller $l_0$ value corresponds to a larger angular size. In contrast, IGE- and IME-dominated substructures have the same typical size for near-$\mathrm{M_{ch}}$ models.

Why does this dichotomy arise? The answer is rooted in the abundance of elements in the ejecta, which is plotted in Fig.~\ref{fig:model_abundance_profiles} as a function of the mass coordinate. We see that the outer layers of the sub-$\mathrm{M_{ch}}$ models in our sample are IGE-poor in comparison to their near-$\mathrm{M_{ch}}$ counterparts. This means the turbulent shocked region in the sub-$\mathrm{M_{ch}}$ models at early times are also IGE-poor (but not IME-poor). RTI substructures formed at early times have a smaller typical size (that is, a larger peak wavenumber $l_0$; see Fig.~\ref{fig:density_ps_dd}). For the sub-$\mathrm{M_{ch}}$ models, these substructures are thus IME-rich but IGE-deficient. The reverse shock encounters more IGEs as it reaches deeper inside the ejecta at late times. By then, the density scale height has increased and allows for formation of typically larger substructures, which are dominated by IGEs. In contrast, the outer layers of the near-$\mathrm{M_{ch}}$ models in our sample are relatively more IGE-rich, exhibiting less ejecta stratification compared to the sub-$\mathrm{M_{ch}}$ models. In these models, IMEs and IGEs are thus similarly involved in the formation of RTI substructures at all times.

The $l_0^{\mathrm{Si}}/l_0^{\mathrm{Fe}}$ ratio for Tycho's SNR, as shown in Fig.~\ref{fig:l0_ratio_ev}, therefore implies significant IGE-deficiency in the outer layers of Tycho's progenitor. This holds true even if the dynamical age of Tycho is considered to be somewhere in the range $t_D=0.4-0.8$ (see section~\ref{subsec:remnant}), instead of taking $t_D=0.6$ (as shown by the analysis in section~\ref{subsec:tycho_ps}). In this work, such radial stratification of IGEs seems to be an exclusive feature of only the sub-$\mathrm{M_{ch}}$ models. Whether or not this property can be robustly attributed to sub-$\mathrm{M_{ch}}$ Type-Ia SNe in general, remains an open question, which will be discussed further in section~\ref{subsec:phys_implications}.

\subsection{Mass ratios in the shocked ejecta}
\label{subsec:mass_ratio}

The deduction that Tycho's SNR likely has a sub-$\mathrm{M_{ch}}$ WD progenitor is based on the dichotomy between the near-$\mathrm{M_{ch}}$ and the sub-$\mathrm{M_{ch}}$ SNR models as found in section~\ref{subsec:PS_of_elements} and~\ref{subsec:l0_ratio}. But the direct comparison of substructures in Tycho's SNR to those in our models also relies on our computation of synthetic narrowband X-ray images, as mentioned in section~\ref{subsec:tycho_ps}. We compute these images assuming that shocked plasma is X-ray bright only when the ionization age $\tau$ is greater than $10^{12}\,\mathrm{cm^{-3}\,s}$. This criterion, along with the ejecta composition, governs the amount of X-ray emitting elements in the shocked ejecta. This in turn determines the substructure distribution of various elements in our synthetic images.

\begin{figure}
\centering
\includegraphics[width=0.48\textwidth]{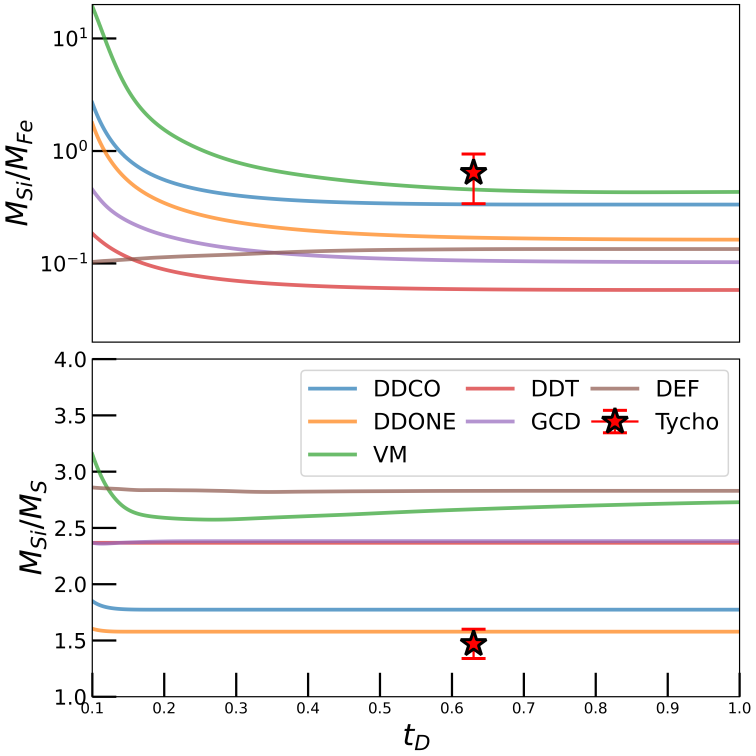}
\vspace{-5mm}
\caption{Time evolution of the mass ratios Si/Fe and Si/S in the shocked ejecta of our near-$\mathrm{M_{ch}}$ (DDT, DEF, and GCD) and sub-$\mathrm{M_{ch}}$ (DDCO, DDONe, and VM) models. The plot also shows Si/Fe and Si/S obtained for Tycho's SNR \citep{HA+2025ApJ} at the dynamical age $t_D\approx0.6$ deduced for Tycho (in section~\ref{subsec:tycho_ps}). The DDCO model, which is our favored model for Tycho (see section~\ref{subsec:tycho_ps}) has mass ratios that are largely consistent with those of Tycho. This serves as a sanity check for our models and our method of computing synthetic X-ray images (see section~\ref{subsec:mass_ratio} for details).
}
\label{fig:mass_ratio_ev}
\end{figure}

Hence, it is important to verify that our simple cutoff criterion makes an estimate of the amount of X-ray bright elements (Si, S and Fe in this case) in our SNR models that is consistent with observed SNRs. We therefore compare the mass ratios of X-ray emitting elements in our hydrodynamic models to those estimated  for Tycho's SNR. Interestingly, these mass ratios have been used to choose the Type-Ia explosion model that is most applicable to Tycho \citep{Badenes+2008ApJ,Katsuda+2015ApJ,MR+2018ApJ,HA+2025ApJ}. In this work, we use the mass ratios of X-ray emitting Si/Fe and Si/S as measured for Tycho by \cite{HA+2025ApJ}. These are shown in Fig.~\ref{fig:mass_ratio_ev}, along with the same for all our models as a function of time. There is no strong dichotomy between the near-$\mathrm{M_{ch}}$ and the sub-$\mathrm{M_{ch}}$ SNR models for either ratio, unlike the case for the $l_0^{\mathrm{Si}}/l_0^{\mathrm{Fe}}$ ratio. This supports the conclusion by \cite{HA+2025ApJ} that  mass ratios alone cannot distinguish near-$\mathrm{M_{ch}}$ from sub-$\mathrm{M_{ch}}$ SNR models. Nevertheless, we find that both the Si/Fe and the Si/S mass ratios for Tycho's SNR are largely consistent with those of the DDCO model, which is our favored model for Tycho. This agreement is not enough in itself to support our result that the DDCO model is the most ideal candidate for Tycho's SNR, but shows that this result is self-consistent. We further caution the reader that the comparison of the mass ratios in our models to those of Tycho only serves as a sanity check and not as a test of accuracy for any particular model. Firstly, these models are representative of different classes of explosion mechanisms and do not systemically cover a range of parameter values that would be relevant for Tycho. Secondly, X-ray emission in the shocked ejecta of SNRs is not governed only by a simple ionization age threshold as implemented in this work, but also by other poorly constrained physical parameters such as the efficiency of collisionless electron heating in the shocks \citep{Badenes+2005ApJ,Yamaguchi+2014ApJ}, or certain nuclear reaction rates \citep[e.g. $^{12}$C+$^{16}$O; see][]{Bravo+2019MNRAS,HA+2025ApJ}. We encourage future studies to further explore the role of different explosion models and detailed NEI calculations on X-ray emission and emissivity distribution in Type-Ia SNRs.

\section{Discussion}    \label{sec:discussion}

\subsection{Physical implication and caveats}
\label{subsec:phys_implications}

In section~\ref{subsec:l0_ratio}, we show that amongst the different SNR models used in this work, the Si and Fe dominated substructure sizes in Tycho's SNR are the most consistent with those in the DDCO model. The substructure size ratio disparity between the near-$\mathrm{M_{ch}}$ and the sub-$\mathrm{M_{ch}}$ SNR models arises from the lack of IGEs in the outer layers of the latter. The fact that Tycho's SNR is found to behave like our sub-$\mathrm{M_{ch}}$ models in this regard implies that SN ejecta remain somewhat stratified, at least in some Type-Ia SNRs. This has been shown to be the case for several Type-Ia SNe, by deriving the abundance profiles of SN nucleosynthesis products in the ejecta using spectral analysis  \citep{Stehle+2005MNRAS,Mazzali+2008MNRAS,Tanaka+2011MNRAS,Sasdelli+2014MNRAS,Ashall+2016MNRAS,Aouad+2022MNRAS}. Our work thus adds an independent, morphological line of evidence supporting this picture.

But this raises a deeper question, as noted in section~\ref{subsec:l0_ratio}: can the observed abundance stratification (or lack thereof) in Type Ia SNe be robustly linked to the explosion channel? Multidimensional near-$\mathrm{M_{ch}}$ SN models are known to exhibit reduced radial stratification, due to buoyant deflagration plumes driving strong turbulent mixing \citep{Seitenzahl+2013MNRAS,Pakmor+2024AandA}. Even though the N1 DDT model considered here is affected relatively less by this effect \citep[see Fig. 8 of][]{Seitenzahl+2013MNRAS}, it is unclear whether such mixing is essential for producing $l_0^{\mathrm{Si}}/l_0^{\mathrm{Fe}} \approx 1$. To test this, we consider 1D near-$\mathrm{M_{ch}}$ SN models as initial conditions for our SNR modeling. By construction, these SN models do not include turbulent mixing, yet they reproduce many observed properties of Type Ia SNe and their remnants \citep{Badenes+2006ApJ,MR+2018ApJ}, and in some cases yield abundance stratification consistent with that inferred from spectral modeling (see previous paragraph).

\begin{figure*}
\centering
\includegraphics[width=0.99\textwidth]{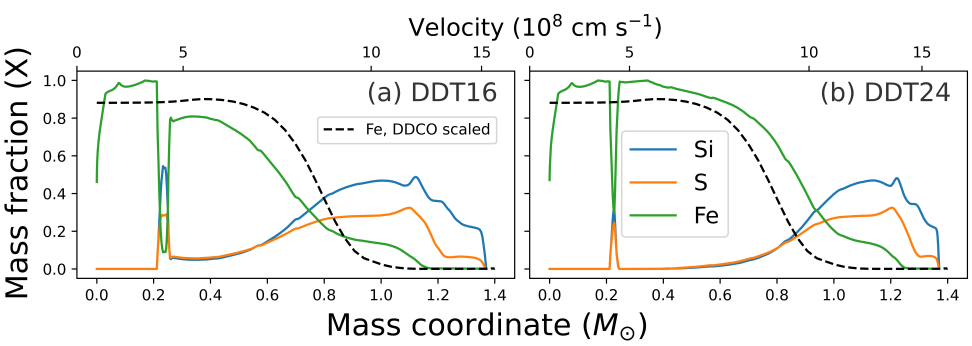}
\vspace{-2mm}
\caption{Mass fraction profiles of Si, S, and Fe as a function of enclosed mass for the DDT16 (\textit{left panel}) and DDT24 (\textit{right panel}) models at solar metallicity \citep{Bravo+2019MNRAS}. For comparison, the Fe abundance profile of the sub-$\mathrm{M_{ch}}$ DDCO model is shown as a dashed line in each panel, rescaled to a total mass of $1.4,\mathrm{M_\odot}$ in order to facilitate a direct comparison of the relative radial distribution of IGEs. Relative to the DDCO model, both DDT16 and DDT24 exhibit enhanced Fe abundances at intermediate mass coordinates. The top axis shows the corresponding velocity to facilitate comparison with other studies.}
\label{fig:1d_ddt_abun_profiles}
\end{figure*}


We therefore compute two additional SNR models, starting from the 1D DDT16 and DDT24 models (at solar metallicity) from \cite{Bravo+2019MNRAS}. The two models were chosen to probe variations in central density of the WD and $^{56}$Ni yield. The DDT24 model, in particular, has been shown to reproduce several observed properties of Tycho’s SNR, including its radius, Fe K$\alpha$ luminosity, and Fe K$\alpha$ centroid energy \citep{MR+2018ApJ}. This provides a controlled setup to isolate the role of radial abundance structure independent of multidimensional mixing. Fig.~\ref{fig:1d_ddt_abun_profiles} shows the corresponding abundance profiles of IMEs and IGEs. Compared to the multi-D DDT model considered earlier, the 1D DDT16 and DDT24 models exhibit reduced IGE abundances at large radii, but still retain more IGEs at intermediate layers than the sub-$\mathrm{M_{ch}}$ DDCO model.

\begin{figure*}
\centering
\includegraphics[width=0.99\textwidth]{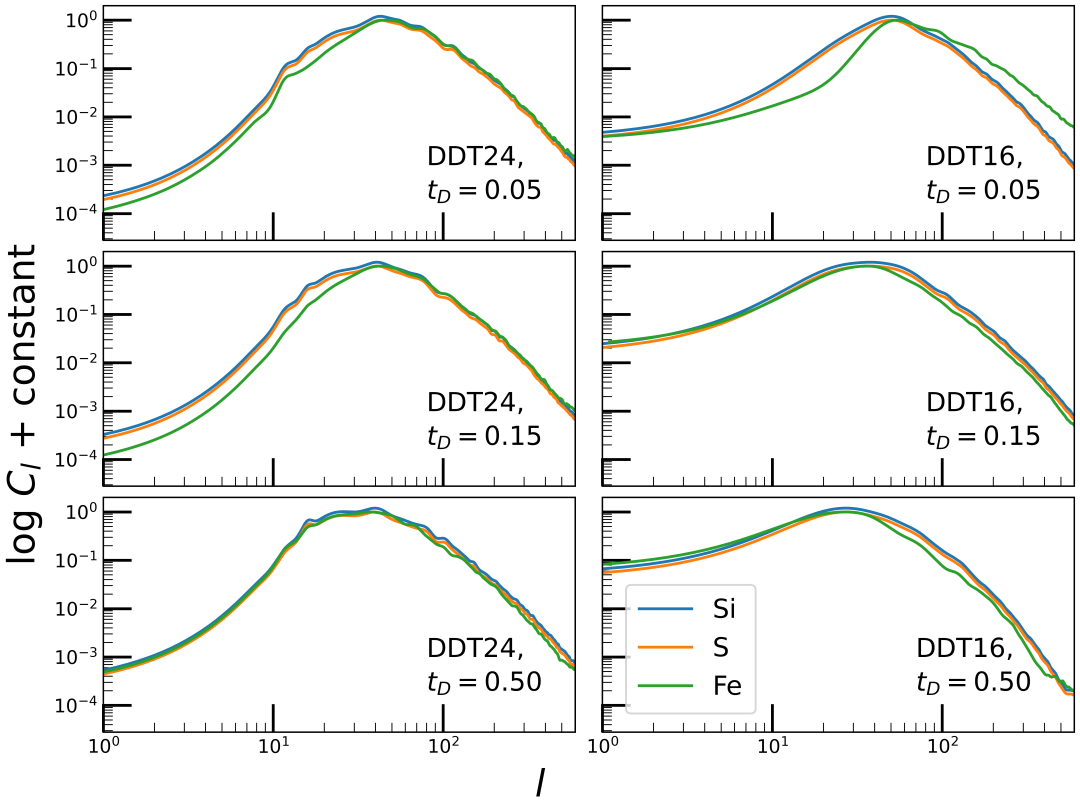}
\vspace{-4mm}
\caption{Time evolution of the power spectra of Si, S and Fe distributions in two SNR models, which use the 1D DDT24 and DDT16 models by \cite{Bravo+2019MNRAS} as their initial conditions. These models are used to test if the abundance stratification in 1D near-$\mathrm{M_{ch}}$ SN models is sufficient to reproduce the size difference in substructures dominated by IMEs and those dominated by IGEs seen in our sub-$\mathrm{M_{ch}}$ SNR models. All the power spectra of the DDT24 and the DDT16 SNR models peak at similar wavenumbers at all instants of time shown here, showing our conclusion (see Fig.~\ref{fig:ps_of_elements}) holds true even if 1D DDT models with very little mixing are considered.}
\label{fig:1D_model_ps}
\end{figure*}

In Fig.~\ref{fig:1D_model_ps}, we show the power spectra of Si, S and Fe for both SNR models at multiple epochs. All power spectra peak at very similar wavenumbers, similar to the near-$\mathrm{M_{ch}}$ SNR models considered earlier, which were developed from spherical averages of 3D SN models. As shown in Fig.~\ref{fig:1d_ddt_abun_profiles}, the 1D DDT models retain non-negligible IGE abundances at intermediate mass coordinates, in contrast to the sub-$\mathrm{M_{ch}}$ DDCO model. This difference alone is sufficient to suppress the scale disparity between Si- and Fe-dominated substructures, yielding $l_0^{\mathrm{Si}}/l_0^{\mathrm{Fe}} \approx 1$ even in the absence of multidimensional mixing.

Our results show that turbulent mixing, as expected in multidimensional near-$\mathrm{M_{ch}}$ SN models, is a contributing but likely not the necessary condition for producing comparable characteristic scales in IME- and IGE-dominated structures. We thus emphasize that the key factor controlling the relation of typical IME- and IGE-dominated substructure sizes is the presence of IGEs in intermediate or outer layers of Type-Ia SN progenitors. In near-$\mathrm{M_{ch}}$ models, the location of IGEs in the ejecta depends sensitively on the density and location at which the deflagration flame transitions to a detonation. In particular, IGEs are synthesized during the detonation phase only when the burning front propagates through material at sufficiently high densities \citep[of order a few $\times 10^7\,\mathrm{g\,cm^{-3}}$;][]{Seitenzahl+2017hsn}. The radial extent of IGE-rich material therefore depends on the degree of pre-expansion prior to detonation and on the adopted transition conditions.

A number of delayed detonation models, both 1D \citep{Iwamoto+1999ApJS} and multi-D \citep{Maeda+2010ApJ,Seitenzahl+2013MNRAS}, do exhibit non-negligible abundance of IGEs in intermediate ejecta layers. As demonstrated by the DDT16/DDT24 models considered here, even a modest presence of IGEs in these layers is sufficient to reduce the scale separation between IME- and IGE-dominated substructures, yielding $l_0^{\mathrm{Si}}/l_0^{\mathrm{Fe}} \approx 1$. Whether such degree of radial IGE mixing is a robust generic outcome across all full near-$\mathrm{M_{ch}}$ WD explosion models, or whether a subset of models can achieve significantly stronger confinement of IGEs to the innermost ejecta (more akin to the sub-$\mathrm{M_{ch}}$ double detonation models), remains an open question. Addressing this will require systematic exploration of the ignition geometry and deflagration-to-detonation transition conditions in delayd detonation models.

It is also important to note that the models used in this work use spherically averaged density and abundance profiles of 3D SN models as their initial conditions. It could be the case that some multi-D near-$\mathrm{M_{ch}}$ SN models achieve a much higher degree of ejecta stratification in some spatial directions compared to others. For example, the DDT models in \cite{Seitenzahl+2013MNRAS} with 3-5 central ignition points exhibit marked asymmetries in element distributions (see their Figs. 2-4). It is possible (but unlikely since SNRs are optically thin) that Tycho's SNR resulted from such an explosion and has high abundance stratification on its observed side. Detailed modeling of SNRs starting from 3D Type-Ia SN models are required to conclusively address this question.

\subsection{Connection to previous studies}
\label{subsec:other_works}

Previous numerical studies have also demonstrated certain differences in the remnant morphology of Type-Ia SNe generated by different explosion mechanisms. For instance, it has been shown \citep{Ferrand+2019ApJ,Ferrand+2021ApJ,Ferrand+2022ApJ} that delayed detonation, double detonation, and pure deflagration mechanisms produce SNRs with qualitatively different large-scale structures. They show that large scale anisotropies in the remnant come predominantly from the SN explosion itself, while small scale anisotropies (including typical size of the substructures) result from RTI activity. Our models cannot probe such large-scale asymmetries (such as a dipolar structure), having started from 1D initial conditions.

Lastly, we note that \cite{Lopez+2011ApJ} have done a closely related study on SNRs, as mentioned in Section \ref{sec:intro}. They calculate the X-ray substructure size distribution at various narrowbands (each corresponding to a different element) for a sample of 24 SNRs. The wavelet transform analysis technique they use \citep{Lopez+2009ApJ} for analysing the SNR images is very similar to the $\Delta$-variance technique used in this paper, but functionally differs by a factor of $l^2$ \citep[][see Section 5]{Mandal+2024ApJ}. This difference arises because these two techniques measure slightly different quantities, and therefore the plots of \cite{Lopez+2011ApJ} cannot be interpreted directly using our results. Regardless, their measurements provide a wealth of information on nearby SNRs and can be used to diagnose explosion mechanisms of the Type-Ia subset, although we suggest exercising caution against direct visual interpretation for the reason discussed above. Moreover, the difference in substructure sizes of different elements is also apparent for several core-collapse SNRs in their sample, such as Cas A and G292.0+1.8. Like the Type-Ia case, this difference is an indicator that different elements are most abundant at different depths in the ejecta. Therefore, they are drawn by the reverse shock into the turbulent shocked zone at different times, making each element reside predominantly in RTI clumps that are not of the same typical size. This is a powerful diagnostic of the SN explosion and can be used to favor certain classes of core-collapse models, like the Type-Ia case.

\subsection{The SN-SNR connection}

While SNRs provide us an extended view of the stellar ejecta (and information on its geometry, kinematics, and chemical abundances), much of the vital information about SN explosions are contained in their spectra and lightcurves. Although this information is not readily available for historical SNe that are now visible as remnants, light echoes of some historical Galactic SNe from neighboring interstellar dust has revealed their spectra and lightcurves, e.g., Tycho's SNR \citep{Krause+2008Nature}, SNR 0509-67.5 \citep{Rest+2008ApJ_0509}, Cassiopeia A \citep{Rest+2008ApJ_CasA}. This provides an exciting opportunity for the present work to be used as a bridge between observed properties of Type-Ia SNe and their explosion mechanism, as revealed by the SNR.

Particularly, in the case of Tycho's SNR, the associated SN (SN 1572) was found to be a normal\footnote{as opposed to overluminous 1991T-like or subluminous \\ 1991bg-like SNe} Type-Ia SN from the optical light-echo spectrum of the SN \citep{Krause+2008Nature} and the X-ray spectrum of the SNR \citep{Badenes+2006ApJ}, as well as the historical lightcurve of the SN \citep{RL2004ApJ}. Perhaps the most unusual feature in the optical spectrum of SN 1572 is the presence of high-velocity Ca II absorption features. Unlike other normal Type-Ia SNe, which have a Ca II photospheric absorption feature corresponding to ${\sim}13{,}000\mathrm{\,km s^{-1}}$, the Ca II absorption lines in SN 1572 correspond to $\geq20{,}000\mathrm{\,km s^{-1}}$. To our knowledge, such a high velocity component has only been detected rarely in type-Ia spectra, for instance in SN 2001el \citep{Mattila+2005A&A}. For SN 2001el, the high-velocity Ca II feature has been shown to be the result of an aspherical explosion using spectropolarimetry \citep{Wang+2003ApJ,Kasen+2003ApJ}. \cite{Krause+2008Nature} speculate that obtaining more light echo spectra of SN 1572 in different spatial directions would shed light on how anisotropic this Ca II feature is, such as whether it is due to accretion from a companion or due to an aspherical explosion mode. Such investigations have already been performed for a galactic SNR \citep[Cas A;][]{Krause+2008Science, Rest+2008ApJ_CasA,Rest+2011ApJ,Besel+2012A&A} and can probe spectroscopic diversity in Type-Ia SNe using a single SNR.

\section{Conclusion}    \label{sec:conclusion}

We demonstrate a way to diagnose the SN explosion mechanism for remnants of Type-Ia or thermonuclear SNe. This technique relies on the previously discovered fact that SN ejecta forms small-scale substructures during its evolution into a remnant, and the typical size of these substructures at a given time is governed by the density scale height of the outermost ejecta \citep{Polin+2022ApJ,Mandal+2023ApJ,Mandal+2024ApJ}. In this work, we develop high resolution three-dimensional SNR models using several different Type-Ia SN models as initial conditions. We investigate whether substructures dominated by different elements have the same typical size (or angular wavenumber $l_0$) or not. Our elements of interest are Si, S, and Fe, which are among prime nucleosynthesis products and produce prominent emission lines in X-rays. We also develop corresponding synthetic narrowband X-ray images to quantitatively compare against corresponding images of Tycho's SNR, as observed by the \textit{Chandra} X-ray Observatory. Our key novel findings are as follows:

\begin{enumerate}
    \item SNRs in our model suite that originate from a near-$\mathrm{M_{ch}}$ WD explosions have Si-, S-, and Fe-dominated substructures of similar sizes, indicating even mixing of elements in the shocked ejecta.
    \item In contrast, sub-$\mathrm{M_{ch}}$ WD explosions (double detonations or WD-WD mergerss) produce Fe-dominated substructures that are significantly larger than Si- or S-dominated substructures. This indicates that for such explosions, IGEs preferentially reside in RTI structures formed in regions of larger density scale height (or shallower density profiles). Such regions are expected to be closer to the center of the remnant.
    \item The power spectra of our models have a steep slope at small scales or large $l$ ($C_l \propto l^{-3}$), indicating a suppression of turbulent cascade in the SNR. During later stages of evolution, the power spectra attain a shallower shape that approaches the Kolmogorov cascade ($C_l \propto l^{-5/3}$). This is similar to the findings of \cite{Mandal+2024ApJ}, who observe this for an exponential density profile ejecta.
    \item For Tycho's SNR, $l_0^{\mathrm{Si}}\approx l_0^{\mathrm{S}}\approx23$, while $l_0^{\mathrm{Fe}}\approx13$. In other words, Fe-dominated substructures are about twice as big as Si- or S-dominated substructures. Moreover, power spectrum for the Fe-rich part of the ejecta is consistent with a shallower slope ($C_l \propto l^{-1.8}$) at small scales than its Si or S counterpart ($C_l \propto l^{-3}$). These facts indicate that most of the IGEs in the turbulent, shocked ejecta come from lower radii in the SN ejecta unlike Si or S, which do not preferentially enter the shocked zone from deeper inside the ejecta. The currently examined models thus favor a sub-$\mathrm{M_{ch}}$ origin for Tycho, but the physical uniqueness of this interpretation relative to near-$\mathrm{M_{ch}}$ models is yet to be demonstrated explicitly.
\end{enumerate}


Although our work focuses on Type-Ia SNRs, we emphasize that this technique provides a general probe of density or abundance distributions of different elements in the ejecta of an SNR. A future extension of this work will be made to diagnose nearby core-collapse SNRs and compare them against existing models.

\acknowledgments
 
We are grateful to the anonymous referee for their thoughtful comments, which significantly improved the manuscript. We thank Dr. Eduardo Bravo for kindly providing the DDT16 and the DDT24 models. This work made use of the Heidelberg Supernova Model Archive (HESMA), https://hesma.h-its.org. Numerical calculations were performed on the Rivanna computing cluster at University of Virginia.

\software{\sprout\, \citep{Mandal+2023_sprout},  
    VisIt \citep{HPV:VisIt}, 
    SHTOOLS \citep{SHTOOLS},
    NumPy \citep{numpy},
    Matplotlib \citep{matplotlib},
    Astropy.
}

\vspace{30mm}
\appendix 

\section{Matching Tycho's data against models}
\label{app:tycho_vs_models}

\begin{figure*}
\centering
\includegraphics[width=0.98\textwidth]{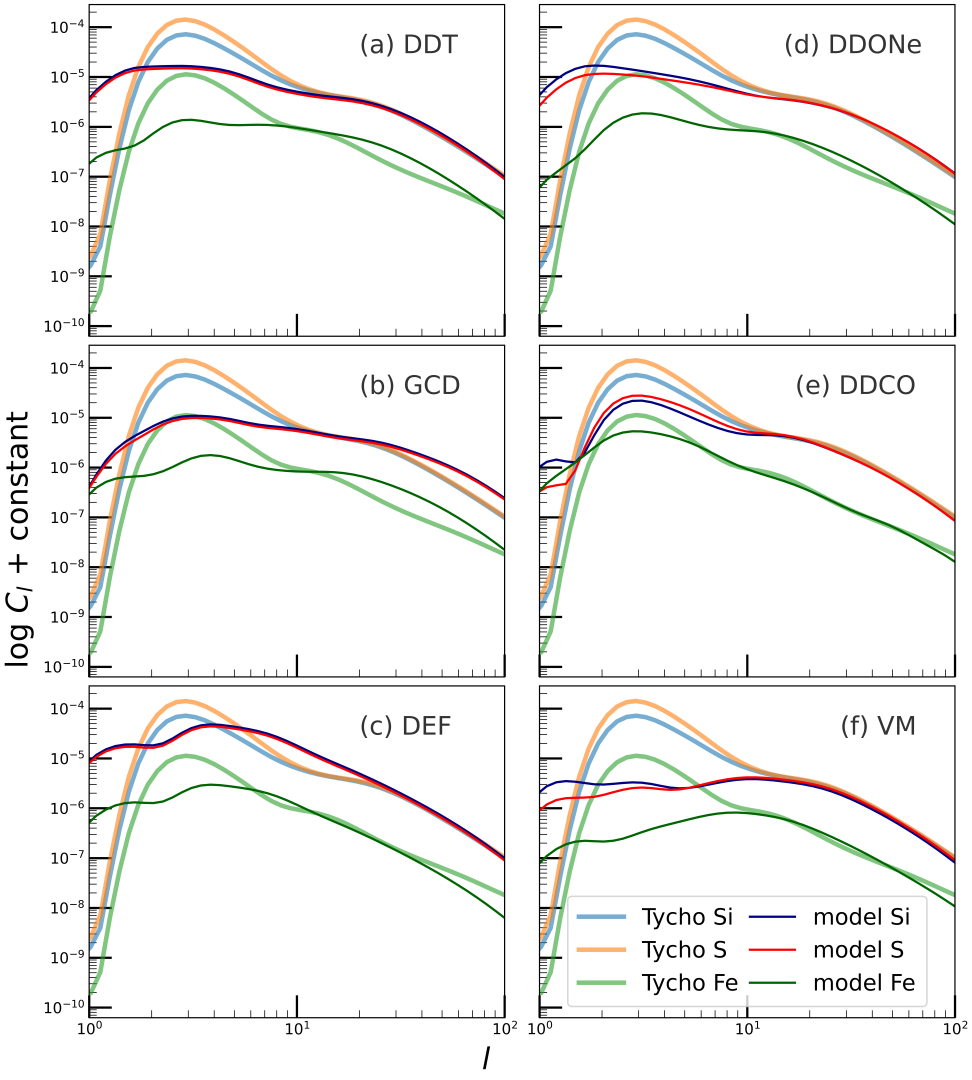}
\vspace{-4mm}
\caption{Power spectra of synthetic narrowband X-ray images (corresponding to Si, S, and Fe) generated from each of our SNR models at $t_D=0.6$. The power spectrum from Tycho's SNR are also overlaid in these plots for comparison. The sub-$\mathrm{M_{ch}}$ models (DDCO, DDONE and VM) have power spectrum that match those of Tycho's SNR the best, with the DDCO model providing the best match. Normalizations of the power spectra were chosen arbitrarily to aid visual inspection.}
\label{fig:models_at_tD_0p6}
\end{figure*}

Here we present comparisons of the power spectrum obtained from the narrowband X-ray images of Tycho's SNR (shown in Fig~\ref{fig:tycho_ps}) to the same obtained from synthetic images of all our models at $t_D=0.6$. These are plotted in Fig~\ref{fig:models_at_tD_0p6}. The DDCO model is found to offer the best match for all three power spectra, even though the VM and the DDONe models are reasonably close. In contrast, the near-$\mathrm{M_{ch}}$ models have difficulty reproducing all three of Tycho's power spectra (corresponding to Si, S and Fe distributions). The Si or S distribution in the DDT model seems to reproduce the $l_0$ value for Tycho reasonably well, but $l_0^{\mathrm{Fe}}$ for the DDT model is substantially bigger than what is observed for Tycho's SNR.

\begin{figure*}
\centering
\includegraphics[width=0.98\textwidth]{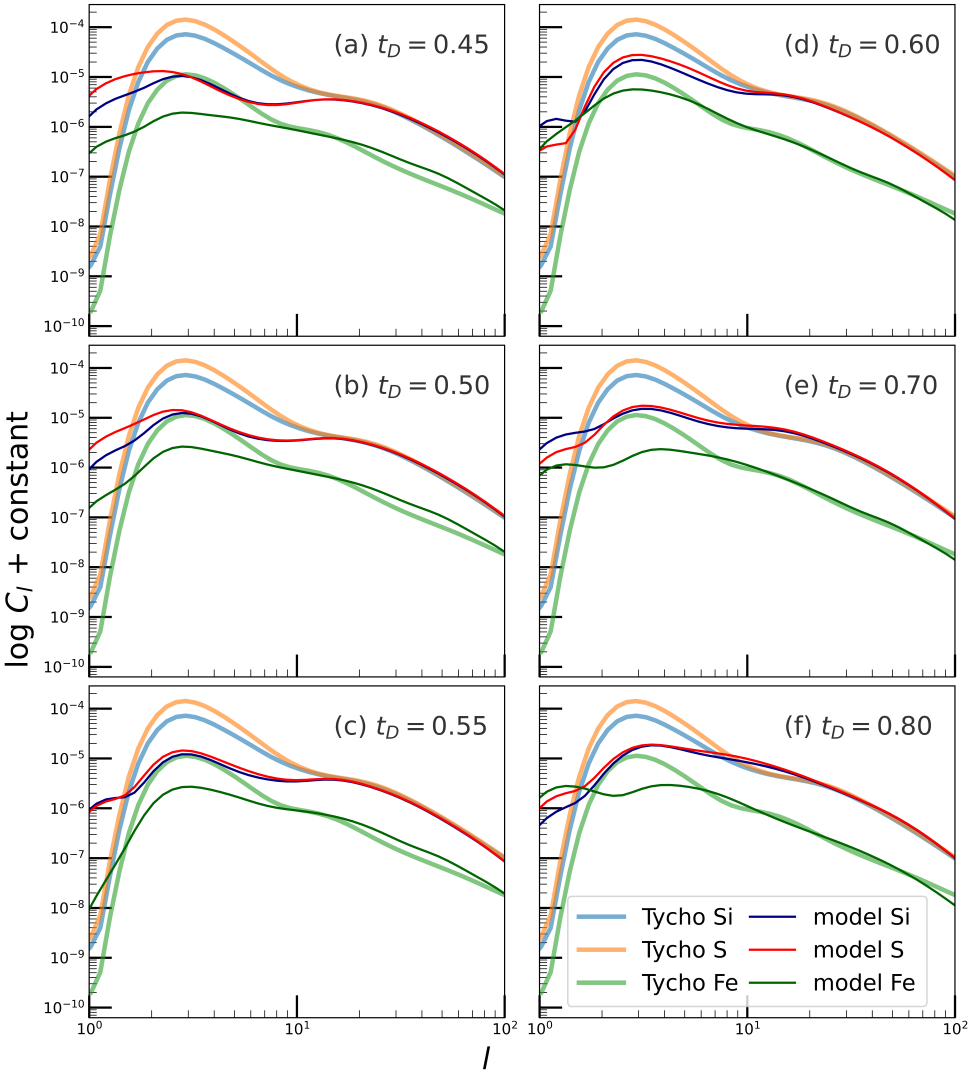}
\vspace{-4mm}
\caption{Power spectra of synthetic narrowband X-ray images (corresponding to Si, S, and Fe) generated from synthetic image of the DDCO models at $t_D=0.45,0.50,0.55,0.70\mathrm{\,and\,}0.90$. The power spectrum from Tycho's SNR are also overlaid in these plots for comparison.}
\label{fig:tycho_vs_ddco}
\end{figure*}

In addition to comparison against all models at $t_D=0.6$, we compare the power spectrum of synthetic images of the DDCO model at time instants $t_D=0.45,0.50,0.55,0.70\mathrm{\,and\,}0.90$, against power spectrum from Tycho's data in Fig~\ref{fig:tycho_vs_ddco}. The power spectrum at $t_D=0.6$ are found to fit Tycho's power spectrum best. For $t_D<0.6$, the power spectrum of Si and S distributions for both Tycho and the DDCO synthetic images match reasonably well at $l>10$. The power spectrum of Fe distribution in the DDCO model at these time instants, however shows a bend at a larger wavenumber compared to that of Tycho. For $t_D>0.6$, all power spectrum obtained from the DDCO model images show the expected bend at smaller wavenumbers compared to the power spectrum from Tycho's SNR.

\section{Synthetic images from all models}
\label{app:all_synthetic_images}

\begin{figure*}
\centering
\includegraphics[width=0.7\textwidth]{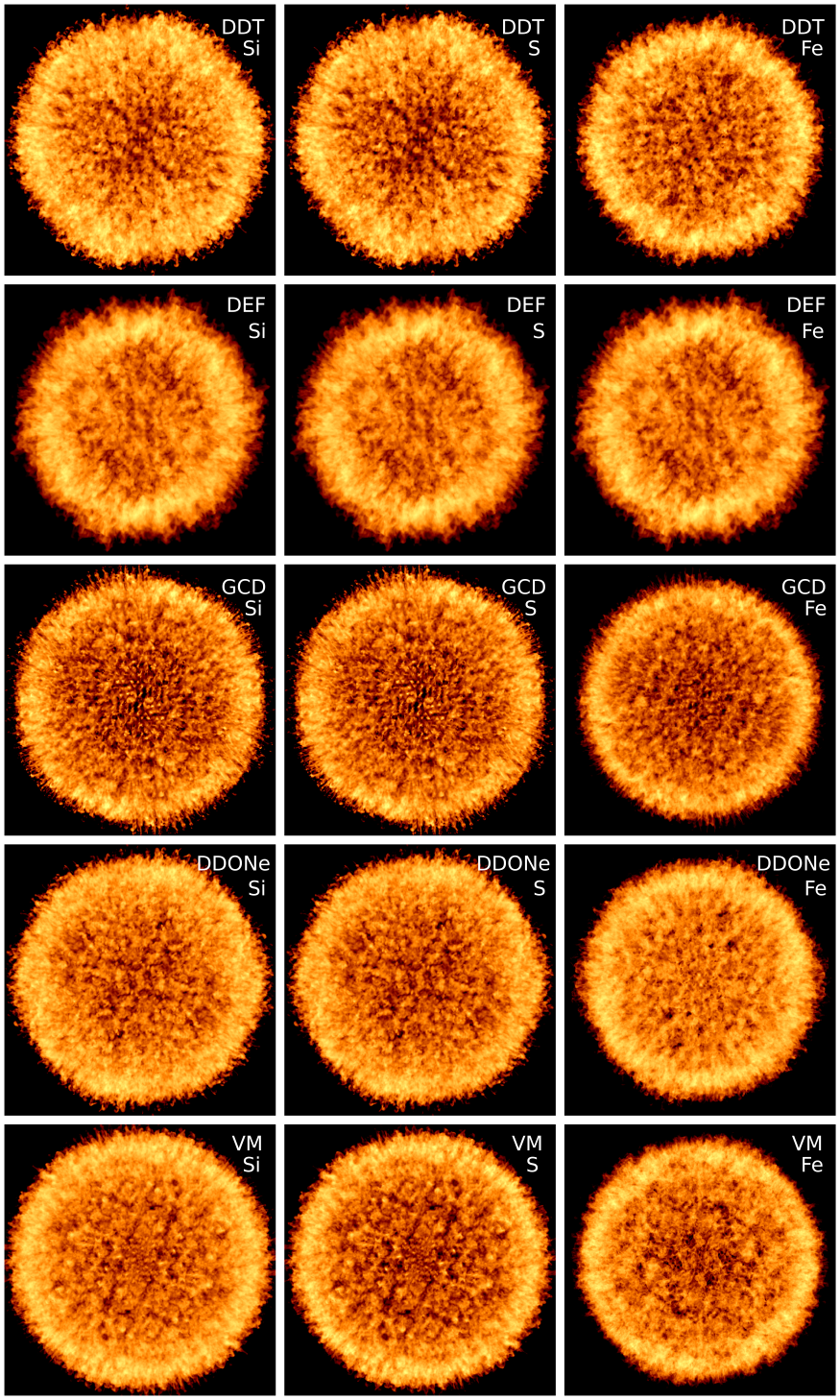}
\vspace{-3mm}
\caption{Synthetic images from all our models at $t_D=0.6$ (except the DDCO model, which can be found in Fig.~\ref{fig:xray_images}). This time instant corresponds to the likely dynamical age of Tycho's SNR. From top to bottom, the rows represent the DDT, DEF, GCD, DDONe and VM models, respectively. From left to right, the columns represent the Si, S, and Fe narrowband images, respectively.}
\label{fig:all_synth_images}
\end{figure*}

To enable visual comparisons of the structural distributions and power spectra, all the two dimensional synthetic images from our SNR models at $t_D=0.6$ are presented in this paper. As we show in section~\ref{subsec:tycho_ps}, the DDCO model at this time instant provides the best match for the morphological properties of Tycho's SNR. Images for the DDCO model at $t_D=0.6$ have been presented in Fig.~\ref{fig:xray_images}. Images for the remaining models at the same time instant are presented in Fig.~\ref{fig:all_synth_images}.

\bibliographystyle{apj} 
\typeout{}
\bibliography{smbib}

\end{document}